\documentclass[12pt,draftclsnofoot,journal,onecolumn]{IEEEtran}

\usepackage{graphicx}
\usepackage[cmex10]{amsmath}
\usepackage{amsfonts}
\usepackage{amssymb}
\usepackage{mathrsfs}
\usepackage{bm}
\usepackage{algorithmic}
\usepackage{algorithm}
\usepackage{array}
\usepackage{mdwmath}
\usepackage{mdwtab}
\usepackage[caption=false,font=footnotesize]{subfig}
\usepackage{fixltx2e}
\usepackage{stfloats}
\usepackage{footnote}
\usepackage{tabularx}
\usepackage{multirow}
\usepackage{color}
\usepackage{textcomp}
\usepackage{subfig}

\newtheorem{definition}{Definition}

\newtheorem{theorem}{Theorem}
\newtheorem{lemma}{Lemma}

\newtheorem{remark}{Remark}

\begin{document}

\title{\huge Energy-efficient Nonstationary Spectrum Sharing}

\author{Yuanzhang~Xiao and~Mihaela~van~der~Schaar,~\IEEEmembership{Fellow,~IEEE} \\
Department of Electrical Engineering, UCLA,
Los Angeles, California 90095. \\
Email: \{yxiao,mihaela\}@ee.ucla.edu.}

\maketitle

\begin{abstract}
We develop a novel design framework for energy-efficient spectrum sharing among autonomous users who aim to minimize their energy consumptions
subject to minimum throughput requirements. Most existing works proposed \emph{stationary} spectrum sharing policies, in which users transmit at
\emph{fixed} power levels. Since users transmit simultaneously under stationary policies, to fulfill minimum throughput requirements, they need to
transmit at high power levels to overcome interference. To improve energy efficiency, we construct \emph{nonstationary} spectrum sharing policies, in
which the users transmit at \emph{time-varying} power levels. Specifically, we focus on TDMA (time-division multiple access) policies in which one
user transmits at each time (but not in a round-robin fashion). The proposed policy can be implemented by each user running a low-complexity
algorithm in a decentralized manner. It achieves high energy efficiency even when the users have \emph{erroneous} and \emph{binary} feedback about
their interference levels. Moreover, it can adapt to the \emph{dynamic entry and exit} of users. The proposed policy is also deviation-proof, namely
autonomous users will find it in their self-interests to follow it. Compared to existing policies, the proposed policy can achieve an energy saving
of up to 90\% when the number of users is high.
\end{abstract}


\section{Introduction}
A key challenge in wireless networks is determining efficient solutions for the autonomous users to share the spectrum. In cognitive radio networks
where the users are differentiated as primary users (PUs) and secondary users (SUs), we also require SUs to access the spectrum without degrading
PUs' quality of service (QoS) \cite{ZhaoSadler}\cite{MusavianAissa_TWC2009}--\cite{MakkiEriksson_TWC2012}. To be more general, we consider cognitive
radio networks in this work, and design spectrum sharing policies that achieve efficient spectrum usage and protect PUs' QoS. Our work can be easily
applied to a wireless network in which users are not differentiated as PUs and SUs (which can be considered as a special cognitive radio network with
no PUs).

Spectrum sharing policies, which specify the PUs' and SUs' transmission schedules and transmit power levels, are essential to achieve spectrum and
energy efficiency \cite{Chiang_FoundationTrend08}. Research on designing spectrum sharing policies can be roughly divided in two main categories. The
research in the first category formulates the spectrum sharing problem as a utility maximization problem subject to the users' maximum transmit power
constraints \cite{TanLow}--\cite{XieYuJi_Infocom2012}\cite{EtkinTse}--\cite{XiaoMihaela_RepeatedGame}\cite{XiaoMihaela_CognitiveRadio}. Many works in
this category \cite{TanLow}--\cite{XiaovanderSchaar_PowerControl}\cite{EtkinTse}--\cite{XiaoMihaela_RepeatedGame}\cite{XiaoMihaela_CognitiveRadio}
define the utility function as an increasing function of the signal-to-interference-and-noise-ratio (SINR), while neglecting to consider the energy
consumption of the resulting spectrum sharing policies. Some other works in this category \cite{SaraydarMandayam_TCOM02}--\cite{XieYuJi_Infocom2012}
define the utility function as the ratio of throughput to transmit power, in order to maximize the spectrum efficiency per energy consumption.
Research in the second category \cite{Yates95}--\cite{SorooshyariTanChiang} formulates the spectrum sharing problem as an energy consumption
minimization problem subject to the users' minimum throughput requirements. In this formulation, the users' throughput requirements can be explicitly
specified. Hence, the spectrum efficiency is guaranteed with the minimal energy consumption. The work in this paper pertains to this second category
of research works.

One major limitation of existing works in the second category \cite{Yates95}--\cite{SorooshyariTanChiang} is that they restrict attention to a simple
class of spectrum sharing policies that require the users to transmit at \emph{fixed} power levels as long as the environment (e.g. the number of
users, the channel gains) does not change\footnote{Although some spectrum sharing policies \cite{Yates95}--\cite{SorooshyariTanChiang} go through a
transient period of adjusting the power levels before converging to the optimal power levels, the users maintain the fixed power levels after the
convergence.}. We call this class of spectrum sharing policies \emph{stationary}. The stationary policies are not energy efficient, because due to
multi-user interference, the users need to transmit at high power levels to fulfill the minimum throughput constraints. To improve energy efficiency,
we study \emph{nonstationary}\footnote{We use ``nonstationary'', instead of ``dynamic'', to describe the proposed policy, because ``dynamic spectrum
sharing'' has been extensively used to describe general spectrum sharing policies in cognitive radio, where SUs access the channel opportunistically.
In this sense, our policy is dynamic. However, our nonstationary policy is different from other dynamic spectrum sharing policies, in that the power
levels are time-varying.} spectrum sharing policies. Specifically, we focus on TDMA (time-division multiple access) spectrum sharing policies, a
class of nonstationary policies in which the users transmit in a TDMA fashion. TDMA policies can achieve high spectrum efficiency that is not
achievable under stationary policies, and greatly improve the energy efficiency of the stationary policies, because of the following two reasons.
First, there is no multi-user interference in TDMA policies. Second, TDMA policies allow users to adaptively switch between transmission and
dormancy, depending on the average throughput they have achieved, for the purpose of energy saving. Note that in the optimal TDMA policies we
propose, users usually do not transmit in the simple round-robin fashion, because of the heterogeneity in their minimum throughput requirements and
channel conditions (see Section~\ref{sec:Motivation} for a motivating example that shows the sub-optimality of round-robin TDMA policies).

Another limitation of existing works in the second category \cite{Yates95}--\cite{SorooshyariTanChiang} is the assumption that each user's receiver
can perfectly estimate the local interference temperature (i.e. the interference and noise power level), and can accurately feed it back to its
transmitter. However, in practice, users cannot perfectly estimate the interference temperature, and can only send limited (quantized) feedback.

In this paper, we provide a novel design framework to construct nonstationary spectrum sharing policies that achieve PUs' and SUs' minimum throughput
requirements with minimal energy consumptions, even when the users have erroneous and very limited (only binary) feedback about their local
interference temperatures. We first prove a key property of the optimal TDMA spectrum sharing policy: each user should choose the same power level
whenever it transmits. This property enables us to solve the policy design problem in two tractable steps: first determine the optimal power levels
before run-time, and then determine the transmission schedule at run-time. We then propose a low-complexity distributed instantaneous throughput
selection (ITS) algorithm for the users to determine their optimal power levels before run-time, and a low-complexity distributed
longest-distance-first (LDF) scheduling algorithm to determine the transmission schedule at run-time. We prove that both algorithms converge linearly
independent of the number of users (i.e. the distance from the optimal solution decreases exponentially, resulting in a logarithmic convergence
time). The proposed policy can also adapt to the dynamic entry and exit of users without affecting the convergence of existing users. Moreover, it is
deviation-proof, meaning that a user cannot improve its energy efficiency over the proposed policy while still fulfilling the throughput requirement.
In this way, autonomous users will find it in their self-interest to adopt the policy.

The rest of the paper is organized as follows. We give detailed comparisons against existing works in Section~\ref{sec:Related}.
Section~\ref{sec:Model} describes the system model for spectrum sharing. Section~\ref{sec:Motivation} gives a motivating example to show the
performance gain achieved by nonstationary policies and the necessity of deviation-proof policies. We formulate and solve the policy design problem
in Section~\ref{sec:Formulation} and Section~\ref{sec:Design}, respectively.
Simulation results are presented in
Section~\ref{sec:Simulation}. Finally, Section~\ref{sec:Conclusion} concludes the paper.

\section{Related Works}\label{sec:Related}
In this section, we provide a comprehensive comparison between the proposed scheme and existing works. The reader could skip this section and go
directly to the system model, if not interested in the detailed comparisons.

Although only some works \cite{Yates95}--\cite{SorooshyariTanChiang} use the same problem formulation as ours, we compare against a wide range of
related works \cite{TanLow}--\cite{XiaoMihaela_CognitiveRadio} to highlight the technical novelty of our work, and to illustrate that the works
\cite{TanLow}--\cite{XieYuJi_Infocom2012}\cite{EtkinTse}--\cite{XiaoMihaela_CognitiveRadio} proposed under different problem formulations cannot be
adapted to our setting.

\subsection{Stationary Spectrum Sharing Policies}
Table~\ref{table:RelatedWork_Stationary} categorizes existing stationary spectrum sharing policies based on four criteria: whether the policy
considers energy efficiency, whether the policy is deviation-proof (against stationary or nonstationary policies), what are the feedback requirements
and the corresponding overhead, and whether they can accommodate a varying number of users. Throughout this section, the feedback is the information
on interference and noise power levels sent from a user's receiver to its transmitter.

\begin{table}\scriptsize
\renewcommand{\arraystretch}{1.0}
\caption{Comparisons against stationary spectrum sharing policies.} \label{table:RelatedWork_Stationary} \centering
\begin{tabular}{|c|c|c|c|c|}
\hline
 & Energy-efficient & Deviation-proof & Feedback (Overhead) & User number \\
\hline
\cite{TanLow}--\cite{TanChiangSrikant_TON2013} & No & No & Error-free, unquantized (Large) & Fixed \\
\hline
\cite{HuangBerry_06_Auction}\cite{XiaovanderSchaar_PowerControl} & No & Against stationary policies & Error-free, unquantized (Large) & Fixed \\
\hline
\cite{SaraydarMandayam_TCOM02}--\cite{PerlazaTembineLasaulceDebbah} & Yes & Against stationary policies & Error-free, unquantized (Large) & Fixed \\
\hline
\cite{TanPalomarChiang_TON2009}\cite{SorooshyariTanChiang} & Yes & Against stationary policies & Error-free, unquantized (Large) & Varying \\
\hline
\cite{EtkinTse}--\cite{LeTreustLasaulce} & No & Against stationary and nonstationary policies & Error-free, unquantized (Large) & Fixed \\
\hline
Proposed & Yes & Against stationary and nonstationary policies & Erroneous, binary (One-bit) & Varying \\
\hline
\end{tabular}
\end{table}

\subsection{Nonstationary Spectrum Sharing Policies}
\begin{table}\scriptsize
\renewcommand{\arraystretch}{1.0}
\caption{Comparisons against nonstationary spectrum sharing policies.} \label{table:RelatedWork_Nonstationary} \centering
\begin{tabular}{|c|c|c|c|c|c|c|}
\hline
 & Energy-efficient & Power control & Users & Feedback (Overhead) & Deviation-proof & User number \\
\hline
\cite{XiaoMihaela_RepeatedGame} & No & Yes & Heterogeneous & Error-free, unquantized (Large) & Yes & Fixed \\
\hline
\cite{FudenbergLevineMaskin94} & No & Applicable & Heterogeneous & Erroneous, limited (Medium) & Yes & Fixed \\
\hline
\cite{ZhaoTongSwamiChen} & No & No & Homogeneous & Erroneous, binary (One-bit) & No & Fixed \\
\hline
\cite{ChenZhaoSwami} & Yes & No & Homogeneous & Erroneous, binary (One-bit) & No & Fixed \\
\hline
\cite{LiuZhao2010}--\cite{LiuZhao2012} & No & No & Homogeneous & Error-free, binary (One-bit) & No & Fixed \\
\hline
Proposed & Yes & Yes & Heterogeneous & Erroneous, binary (One-bit) & Yes & Varying \\
\hline
\end{tabular}
\end{table}

There have been some works that develop nonstationary policies using repeated games \cite{XiaoMihaela_RepeatedGame}\cite{FudenbergLevineMaskin94},
Markov decision processes (MDPs) \cite{ZhaoTongSwamiChen}\cite{ChenZhaoSwami}, and multi-art bandit \cite{LiuZhao2010}--\cite{LiuZhao2012}. We
summarize the major differences between the existing nonstationary policies and our proposed policy in Table~\ref{table:RelatedWork_Nonstationary}.

\subsection{Comparison With Our Previous Work}
Most related to this work is our previous work \cite{XiaoMihaela_CognitiveRadio}. However, the design frameworks proposed in
\cite{XiaoMihaela_CognitiveRadio} and in this work are significantly different because the design objectives are different. In
\cite{XiaoMihaela_CognitiveRadio}, we aimed to design TDMA spectrum sharing policies that maximize the users' total throughput without considering
energy efficiency. Under this design objective, each user will transmit at the maximum power level in its slot, as long as the interference
temperature constraint is not violated. Hence, what we optimized was \emph{only the transmission schedule of the users}. In this work, since we aim
to minimize the energy consumption subject to the minimum throughput requirements, we need to optimize \emph{both the transmission schedule and the
users' transmit power levels}, which makes the design problem more challenging. Moreover, this work considers the scenario in which users enter and
leave the network, which is not considered in \cite{XiaoMihaela_CognitiveRadio}.

\subsection{Comparison With Theoretical Frameworks}

\begin{table}
\renewcommand{\arraystretch}{1.0}
\caption{Comparisons With Related Theoretical Frameworks.} \label{table:RelatedWork_MathematicalFramework} \centering
\begin{tabular}{|c|c|c|c|c|}
\hline
 & Constructive & Discount factor & Feedback & User number \\
\hline
\cite{APS} & No & fixed, $<1$ & N/A & Fixed \\
\hline
\cite{FudenbergLevineMaskin94} & No & $\rightarrow 1$ & Erroneous, high-granularity & Fixed \\
\hline
Proposed & Yes & fixed, $<1$ & Erroneous, binary & Varying \\
\hline
\end{tabular}
\end{table}

Our results on nonstationary policies build on the concept of ``self-generating sets'' proposed in the game theory literature \cite{APS}.
Self-generating sets are used to analyze repeated games with imperfect monitoring. For example, the Folk Theorem in repeated games with imperfect
monitoring in \cite{FudenbergLevineMaskin94} builds on the concept of self-generating sets. However, we cannot apply this concept straightforwardly
or in a way similar as in \cite{FudenbergLevineMaskin94} for the following reasons. The self-generating set is defined as a fixed point of a
set-valued mapping. The work \cite{APS} defined the set-valued mapping, and proved an important property of the fixed point of this set-valued
mapping (i.e. the self-generating set): every payoff vector in the self-generating set can be achieved at an equilibrium. However, although
\cite{APS} discovered this important property, it did not show how to construct a self-generating set. Without constructing the self-generating set,
we do not know what payoff vectors can be achieved at the equilibria or how to achieve them.

The concept of self-generating sets is applied in \cite{FudenbergLevineMaskin94} to prove the Folk theorem in repeated games with imperfect
monitoring. However, our work is fundamentally different from \cite{FudenbergLevineMaskin94} in two aspects. First, the results in
\cite{FudenbergLevineMaskin94} are not constructive: they focus on what payoff vectors can be achieved, but not how to achieve them. In contrast,
given a target payoff vector, we explicitly construct the policy to achieve it. Second, the results in \cite{FudenbergLevineMaskin94} require a
high-granularity feedback signal, namely the cardinality of feedback signals should be proportional to the number of power levels a user can choose.
In contrast, by exploiting the structure of the spectrum sharing problem, we prove that binary feedback is sufficient to achieve optimality in the
considered scenarios.

In Table~\ref{table:RelatedWork_MathematicalFramework}, we summarize the key differences between our work and
\cite{FudenbergLevineMaskin94}\cite{APS}.

\section{System Model}\label{sec:Model}

\subsection{Model For Spectrum Sharing in Cognitive Radio Networks}
We consider a cognitive radio network that consists of $M$ primary users and $N$ secondary users transmitting in a single frequency channel. The set
of PUs and that of SUs are denoted by $\mathcal{M}\triangleq\{1,2,\ldots,M\}$ and $\mathcal{N} \triangleq \{M+1,M+2,\ldots,M+N\}$, respectively. A
wireless network in which users are not differentiated as PUs and SUs is a special case of our model with $M=0$. Each user\footnote{We refer to a
primary user or a secondary user as a user in general, and will specify the type of users only when necessary.} has a transmitter and a receiver. The
channel gain from user $i$'s transmitter to user $j$'s receiver is $g_{ij}$. Each user $i$ chooses its power level $p_i$ from a compact set
$\mathcal{P}_i\subseteq\mathbb{R}_+$. We assume that $0\in\mathcal{P}_i$, namely user $i$ can choose not to transmit. The set of joint power profiles
is denoted by $\bm{\mathcal{P}}=\prod_{i=1}^{M+N} \mathcal{P}_i$, and the joint power profile of all the users is denoted by
$\bm{p}=(p_1,\ldots,p_{M+N}) \in \bm{\mathcal{P}}$. Let $\bm{p}_{-i}$ be the power profile of all the users other than user $i$. Each user $i$'s
throughput is a function of the joint power profile, namely $r_i:\bm{\mathcal{P}}\rightarrow\mathbb{R}_+$. Since the users cannot jointly decode
their signals, each user $i$ treats the interference from the other users as noise, and obtains the following throughput at the power profile
$\bm{p}$ \cite{Chiang_FoundationTrend08}--\cite{LeTreustLasaulce}:
\begin{eqnarray}\label{eqn:Throughput}
r_i(\bm{p}) = \log_2\left(1+\frac{p_i g_{ii}}{\sum_{j\in\mathcal{M}\cup\mathcal{N},j\neq i} p_j g_{ji} + \sigma_i^2}\right).
\end{eqnarray}
where $\sigma_i^2$ is the noise power at user $i$'s receiver.

We define user $i$'s local interference temperature $I_i(\bm{p}_{-i})$ as the interference and noise power level at its receiver, namely
$I_i(\bm{p}_{-i})\triangleq\sum_{j\in\mathcal{M}\cup\mathcal{N},j\neq i} p_j g_{ji} + \sigma_i^2$. We assume that each user $i$ measures the
interference temperature with errors. The estimate of $I_i$ is $\hat{I}_i \triangleq I_i + \varepsilon_i$, where $\varepsilon_i$ is the additive
estimation error with a probability distribution function $f_{\varepsilon_i}$ known to user $i$. Each user $i$'s receiver quantizes $\hat{I}_i$
before feedback it to the transmitter. The quantization function is written as $Q_i:R\rightarrow \mathcal{Q}_i$ with $\mathcal{Q}_i$ being a finite
set of reconstruction values. Given the estimate $\hat{I}_i$, user $i$'s receiver sends the reconstruction value $Q_i(\hat{I}_i)$ to its transmitter.

In this paper, we assume that each user's receiver uses an unbiased estimator such that
$\mathbb{E}_{\varepsilon_i}\{\hat{I}_i(\bm{p}_{-i})\}=I_i(\bm{p}_{-i})$ for any $\mathbf{p}_{-i}$, where $\mathbb{E}_{\varepsilon_i}\{\cdot\}$ is the
expectation over $\varepsilon_i$, and a simple two-level quantizer that preserves the mean value of $\hat{I}_i(\bm{p}_{-i})$ when there is no
multi-user interference. In other words, when $\bm{p}_{-i}=\bm{0}$ (i.e. $I_i(\bm{p}_{-i})=\sigma_i^2$), the quantizer should satisfy
$\mathbb{E}_{\varepsilon_i}\{Q_i(\hat{I}_i(\bm{p}_{-i})|_{\bm{p}_{-i}=\mathbf{0}})\}=\mathbb{E}_{\varepsilon_i}\{\hat{I}_i(\bm{p}_{-i})|_{\bm{p}_{-i}=\bm{0}}\}$,
and thus satisfy $\mathbb{E}_{\varepsilon_i}\{Q_i(\hat{I}_i(\bm{p}_{-i})|_{\bm{p}_{-i}=\bm{0}})\}=I_i(\bm{0})=\sigma_i^2$. An example two-level
quantizer that meets the requirement can be
\begin{eqnarray}\label{eqn:Quantizer}
Q_i(\hat{I}_i(\bm{p}_{-i})) = \left\{\begin{array}{ll} \bar{I}_i\triangleq\int_{x-\sigma_i^2\in\mathrm{supp}(f_{\varepsilon_i}),~x\geq \theta_i} x \cdot f_{\varepsilon_i}(x-\sigma_i^2) dx, & \!\!\mathrm{if}~\hat{I}_i(\bm{p}_{-i})> \theta_i \\
\underline{I}_i\triangleq\int_{x-\sigma_i^2\in\mathrm{supp}(f_{\varepsilon_i}),~x< \theta_i} x \cdot f_{\varepsilon_i}(x-\sigma_i^2) dx, &
\!\!\mathrm{otherwise}\end{array}\right.\!\!\!\!,\forall \bm{p}_{-i}\in\bm{\mathcal{P}}\setminus\mathcal{P}_i,
\end{eqnarray}
where $\mathrm{supp}(f_{\varepsilon_i})$ is the support of the distribution $f_{\varepsilon_i}$, and $\theta_i$ is the quantization threshold. In
practice, it is easy to implement an unbiased estimator and the two-level quantizer in \eqref{eqn:Quantizer}. As we will show, such an estimator and
a quantizer are sufficient to achieve the optimal performance.

\begin{remark}
Here is an intuition why an unbiased estimator and the two-level quantizer in \eqref{eqn:Quantizer} are good enough for us. For user $i$ to achieve a
minimum throughput $r_i$, given the feedback $Q_i(\hat{I}_i)$, its transmit power level $\hat{p}_i$ should be $\hat{p}_i = (2^{r_i}-1)\cdot
Q_i(\hat{I}_i) / g_{ii}$. In a TDMA policy, there is no multi-user interference (i.e. $\bm{p}_{-i}=\mathbf{0}$) when user $i$ transmits. Hence, using
an unbiased estimator and the quantizer in \eqref{eqn:Quantizer}, user $i$'s expected transmit power level is
\begin{eqnarray}
\mathbb{E}_{\varepsilon_i}\left\{\hat{p}_i\right\}=\mathbb{E}_{\varepsilon_i} \left\{(2^{r_i}-1)\cdot Q(\hat{I}_i) / g_{ii}\right\} = (2^{r_i}-1)
\mathbb{E}_{\varepsilon_i}\{Q(\hat{I}_i)\} / g_{ii} = (2^{r_i}-1) \sigma_i^2 / g_{ii},
\end{eqnarray}
which is exactly the transmit power level when user $i$ perfectly knows the interference temperature $\sigma_i^2$. In contrast, under a non-TDMA
policy, there is multi-user interference. In this case, one user's erroneous and quantized feedback affects its own transmit power level, which in
turn affects the other users' transmit power levels through the interference. Thus, all the users' transmit power levels are coupled through the
interference under estimation and quantization errors. Hence, an unbiased estimator and a simple two-level quantizer in \eqref{eqn:Quantizer} may
result in performance loss under non-TDMA policies.
\end{remark}

Since each user $i$ adopts a two-level quantizer, its feedback from the receiver to the transmitter is binary. Then we can further reduce the
feedback overhead as follows. Each user $i$'s receiver informs its transmitter of the two reconstruction values $\bar{I}_i$ and $\underline{I}_i$
only once, at the beginning, after which the receiver sends a signal, probably in the form of a simple probe, only when the estimated interference
temperature $\hat{I}_i$ exceeds the quantization threshold $\theta_i$. The event of receiving or not receiving the probing signal, which is sent only
when $\hat{I}_i>\theta_i$, is enough to indicate user $i$'s transmitter which one of the two reconstruction values it should choose. Since the
probing signal indicates high interference temperature, we call it the \emph{distress signal} as in \cite{BambosPottie},\cite{SorooshyariTanChiang}.
With some abuse of definition, we denote user $i$'s distress signal as $y_i\in Y=\left\{0,1\right\}$ with $y_i=1$ representing the event that user
$i$'s distress signal is sent (i.e. $\hat{I}_i>\theta_i$). We write $\rho_i(y_i|\mathbf{p})$ as the conditional probability distribution of user
$i$'s distress signal $y_i$ given power profile $\bm{p}$, which is calculated as
\begin{eqnarray}\label{eqn:ConditionalDistribution_PowerControl}
\rho_i(y_i=1|\bm{p}) = \int_{x> \theta_i-I_i(\bm{p}_{-i})} f_{\varepsilon_i}(x) dx,~\mathrm{and}~\rho_i(y_i=0|\bm{p}) = 1-\rho_i(y_i=1|\bm{p}).
\end{eqnarray}

\subsection{Spectrum Sharing Policies}
The system is time slotted at $t=0,1,2,\ldots$. At the beginning of time slot $t$, each user $i$ chooses its transmit power $p_i^t$, and achieves the
throughput $r_i(\bm{p}^t)$. At the end of time slot $t$, each user $j$ who transmits ($p_j^t>0$) sends its distress signal $y_j^t=1$ if the estimate
$\hat{I}_j$ exceeds the threshold $\theta_j$. We define $y\in Y$ as the \emph{system distress signal}, indicating whether there exists a user who has
sent its distress signal, namely $y=1$ if there exists $j$ such that $p_j>0$ and $y_j=1$, and $y=0$ otherwise. The conditional distribution is
denoted $\rho(y|\bm{p})$, which is calculated as $\rho(y=0|\bm{p})=\Pi_{j:p_j>0} \rho_j(y_j=0|bm{p})$. Note that the system distress signal is not a
physical signal sent in the system, but rather a logical signal summarizing the status of the system. From now on, we refer to the system distress
signal simply as the distress signal.

Each user $i$ determines the transmit power level $p_i^t$ based on the history of distress signals. The history of distress signals is
$h^t=\{y^0;\ldots;y^{t-1}\}\in Y^t$ for $t\geq1$, and $h^0=\varnothing$ for $t=0$. Then each user $i$'s strategy $\pi_i$ is a mapping from the set of
all the possible histories to its action set, namely $\pi_i:\cup_{t=0}^\infty Y^t\rightarrow\mathcal{P}_i$. The \emph{spectrum sharing policy},
denoted by $\bm{\pi}=(\pi_1,\ldots,\pi_{M+N})$, is the joint strategy profile of all the users. Hence, user $i$'s transmit power level at time slot
$t$ is determined by $p_i^t=\pi_i(h^t)$, and the users' joint power profile is determined by $\bm{p}^t=\bm{\pi}(h^t)$.

We classify all the spectrum sharing policies into two categories, stationary and nonstationary policies. As in \cite[pp.~22]{Altman} and
\cite[Sec.~5.5.2]{MailathSamuelson}, stationary policies always choose the same action under the same state, while nonstationary policies may choose
different actions under the same state. In our model, the state can be considered as the system parameters (e.g. the number of users, the channel
conditions, etc.). Hence, a spectrum sharing policy $\bm{\pi}$ is \emph{stationary} if and only if for all $i\in\mathcal{N}$, for all $t\geq0$, and
for all $h^t\in Y^t$, we have $\pi_i(h^t)=p_i^{\rm stat}$, where $p_i^{\rm stat}\in\mathcal{P}_i$ is a constant. A spectrum sharing policy is
\emph{nonstationary} if it is not stationary. In this paper, we restrict our attention to a special class of nonstationary polices, namely TDMA
policies (with fixed transmit power levels). A spectrum sharing policy $\bm{\pi}$ is a TDMA policy if at most one user transmits in each time slot.
TDMA policies are optimal when the interference among the users is strong \cite{MakkiEriksson}, which is often the case when the number of users is
large. We will illustrate how TDMA policies outperform stationary policies through a simple example in Section~\ref{sec:Motivation} and through
extensive simulations in Section~\ref{sec:Simulation}.

\begin{remark}
In the formal definition of a nonstationary policy, it seems that each user needs to keep track of the history of all the past distress signals at
each time slot. However, as we will see from the longest-distance-first scheduling algorithm that implements the proposed policy, each user only
needs a finite memory.
\end{remark}

\subsection{Definition of Spectrum and Energy Efficiency}
We characterize the spectrum and energy efficiency of a spectrum sharing policy by the users' \emph{discounted} average throughput and
\emph{discounted} average energy consumption, respectively. Each user discounts its future throughput and energy consumption because of its
\emph{delay-sensitive} application (e.g. video streaming) \cite{EtkinTse}--\cite{XiaoMihaela_RepeatedGame}\cite{XiaoMihaela_CognitiveRadio}. A user
running a more delay-sensitive application discounts more (with a lower discount factor). Assuming as in
\cite{EtkinTse}--\cite{FudenbergLevineMaskin94} that all the users have the same discount factor $\delta\in[0,1)$, user $i$'s average throughput is
\begin{eqnarray}\label{AverageThroughput}
R_i(\bm{\pi}) = (1-\delta) \left[ r_i(\bm{p}^0) + \sum_{t=1}^\infty \delta^t \cdot \!\!\!\! \sum_{y^{t-1}\in Y} \!\!\!\!\rho(y^{t-1}|\bm{p}^{t-1})
r_i(\bm{p}^t)\right], \nonumber
\end{eqnarray}
where $\bm{p}^0$ is determined by $\bm{p}^0=\bm{\pi}(\varnothing)$, and $\bm{p}^t$ for $t\geq1$ is determined by
$\bm{p}^t=\bm{\pi}(h^t)=\bm{\pi}(h^{t-1};y^{t-1})$. Similarly, user $i$'s average energy consumption is the expected discounted average transmit
power per time slot, written as
\begin{eqnarray}\label{AverageTransmitPower}
P_i(\bm{\pi}) = (1-\delta) \left[ p_i^0 + \sum_{t=1}^\infty \delta^t \cdot \!\!\!\!\sum_{y^{t-1}\in Y} \!\!\!\!\rho(y^{t-1}|\bm{p}^{t-1})
p_i^t\right]. \nonumber
\end{eqnarray}

Each user $i$ aims to minimize its average energy consumption $P_i(\bm{\pi})$ while fulfilling a minimum throughput requirement $R_i^{\rm min}$. From
one user's perspective, it has the incentive to deviate from a given spectrum sharing policy, if by doing so it can fulfill the minimum throughput
requirement with a lower average energy consumption. Hence, we can define deviation-proof policies as follows.
\begin{definition}\label{def:DeviationProof}
A spectrum sharing policy $\bm{\pi}$ is deviation-proof if for all $i\in\mathcal{M}\cup\mathcal{N}$, we have
\begin{eqnarray}
\pi_i = \arg\min_{\pi_i^\prime} P_i(\pi_i^\prime,\bm{\pi}_{-i}),~\mathrm{subject~to}~ R_i(\pi_i^\prime,\bm{\pi}_{-i})\geq R_i^{\rm min},
\end{eqnarray}
where $\bm{\pi}_{-i}$ is the joint strategy profile of all the users except user $i$.
\end{definition}


\section{Motivation For Deviation-proof TDMA Policies}\label{sec:Motivation}
Before formally describing the design framework, we provide a motivating example to show the advantage and necessity of deviation-proof TDMA
policies. Consider a simple network with two symmetric SUs. The direct channel gains are both $1$, and the cross channel gains are both $\alpha>0$.
The noise at each user' receiver has the same power $\sigma^2$. Both users' minimum throughput requirements are $r$. We first show that a simple
round-robin TDMA policy is more energy-efficient than the optimal stationary policy, and that the optimal TDMA policy outperforms round-robin TDMA
policies. Finally, we demonstrate the necessity of deviation-proofness.

If the users adopt the stationary spectrum sharing policy, to fulfill minimum throughput requirements, their minimum transmit power should be
$p_1^{\rm stat}=p_2^{\rm stat}=\frac{(2^r-1)}{1-(2^r-1)\alpha} \cdot \sigma^2$. The average energy consumptions are then $P_i^{\rm stat}=p_i^{\rm
stat}, i=1,2$, which increase with the cross interference level $\alpha$. Moreover, the stationary policy is infeasible when $\alpha\geq
\frac{1}{2^r-1}$, namely when the cross interference level $\alpha$ or the minimum throughput requirement $r$ is very high.

Now suppose that the users adopt a simple round-robin TDMA policy, in which user 1 transmits at a fixed power level $p_1^{\rm rr}$ in even time slots
$t=0,2,\ldots$ and user 2 transmits at a fixed power level $p_2^{\rm rr}$ in odd time slots $t=1,3,\ldots$. The users' average throughput are
\begin{eqnarray}
R_1 = (1-\delta)\cdot \sum_{t=0}^\infty \delta^{2t} \log_2\left(1+p_1^{\rm rr}/\sigma^2\right) = \frac{1}{1+\delta} \log_2\left(1+p_1^{\rm rr}/\sigma^2\right), \nonumber \\
R_2 = (1-\delta)\cdot \sum_{t=0}^\infty \delta^{2t+1} \log_2\left(1+p_2^{\rm rr}/\sigma^2\right) = \frac{\delta}{1+\delta} \log_2\left(1+p_2^{\rm
rr}/\sigma^2\right). \nonumber
\end{eqnarray}
Given their minimum throughput requirements $r$, we can calculate $p_1^{\rm rr}$ and $p_2^{\rm rr}$ from the above equations, and obtain their
average energy consumptions as
\begin{eqnarray}
P_1^{\rm rr} = (1-\delta) \sum_{t=0}^\infty \delta^{2t} p_1^{\rm rr} = \frac{\sigma^2}{1+\delta} \left(2^{r(1+\delta)}-1\right), P_2^{\rm rr} =
(1-\delta) \sum_{t=0}^\infty \delta^{2t+1} p_2^{\rm rr} = \frac{\sigma^2\delta}{1+\delta} \left(2^{r(1+\frac{1}{\delta})}-1\right). \nonumber
\end{eqnarray}
Note that, as opposed to the stationary policy, the average transmit power in the round-robin TDMA policy is independent of the cross interference
level. Hence, the round-robin TDMA policy is better under medium to high interference levels, the scenarios in which the stationary policy may not
even be feasible. For example, when $r=1$ and $\delta=0.9$, the round-robin TDMA policy is more energy efficient when $\alpha\geq 0.34$.

Under the same parameters (i.e. $r=1$ and $\delta=0.9$), the optimal TDMA policy that achieves the minimum total average energy consumption is not a
round-robin TDMA policy. The transmission schedule of the first few time slots is ``1221122112\ldots'', which seems to follow an irregular pattern,
instead of a round-robin pattern. We will show how to construct the optimal TDMA policy in Section~\ref{sec:Formulation}, and demonstrate its
performance gains in Section~\ref{sec:Simulation}.

Even if a TDMA policy is already energy-efficient, a user may want to deviate from it to achieve higher energy efficiency.
We derive the conditions under which it is beneficial for a user to deviate from a given policy in the following lemma.
\begin{lemma}\label{lemma:ConditionOfDeviation}
Suppose that under a given TDMA policy, user $i$ transmits at power level $p_i^t$ at time $t$ and user $j$ transmits at power level $p_j^{t+s}$ at
time $t+s$, where $t,t+s\geq0$ and $s\neq0$. Then regardless of the discount factor $\delta$, user $j$ can deviate by transmitting in both time slot
$t$ and $t+s$ to achieve at least the same throughput with a lower average energy consumption, if and only if $p_j^{t+s} g_{jj} > p_i^t g_{ij}$.
\end{lemma}
\begin{IEEEproof}
See \cite[Appendix~A]{Appendix}.
\end{IEEEproof}
From the above lemma, we can see that user $j$ has the incentive to deviate when $g_{ji} p_i^t$ is small, namely the interference from user $i$ is
small, and when $p_j^{t+s}$ is large, namely user $j$'s required throughput is high.

%
%

\section{The Design Problem Formulation}\label{sec:Formulation}
Our goal is to construct a deviation-proof TDMA policy that fulfills all the users' minimum throughput requirements and optimizes a certain energy
efficiency criterion. The energy efficiency criterion can be represented by a function defined on all the users' average energy consumptions,
$E(P_1(\bm{\pi}),\ldots,P_{M+N}(\bm{\pi}))$. Note, importantly, that the energy efficiency criterion can also reflect the priority of the PUs over
the SUs. For example, the energy efficiency criterion can be the weighted sum of all the users' energy consumptions, i.e.
$E(P_1(\bm{\pi}),\ldots,P_{M+N}(\bm{\pi})) = \sum_{i\in\mathcal{M}\cup\mathcal{N}} w_i\cdot P_i(\bm{\pi})$ with $w_i\geq0$ and
$\sum_{i\in\mathcal{M}\cup\mathcal{N}} w_i=1$. Each user $i$'s weight $w_i$ indicates the importance of this user. We can set higher weights for PUs
and lower weights for SUs.

Given each user $i$'s minimum throughput requirement $R_i^{\rm min}$, we can formally define the policy design problem as
\begin{eqnarray}\label{eqn:PolicyDesignProblem}
&\displaystyle\min_{\bm{\pi}}& E(P_1(\bm{\pi}), \ldots, P_{M+N}(\bm{\pi})) \\
&s.t.& \bm{\pi}~\mathrm{is~a~deviation-proof~TDMA~policy}, \nonumber\\
&    & R_i(\bm{\pi})\geq R_i^{\rm min},~\forall i\in\mathcal{M}\cup\mathcal{N}. \nonumber
\end{eqnarray}
In the above problem formulation, the usual constraints on the interferences caused by SUs to PUs are satisfied by restricting to TDMA policies, in
which there is no multi-user interference.

\section{A Design Framework For Spectrum and Energy Efficient Policies}\label{sec:Design}
We first outline the procedure to solve the policy design problem \eqref{eqn:PolicyDesignProblem}. Then we show in detail how to solve the design
problem, and discuss implementation issues. Finally, we adapt the proposed policy to the dynamic entry and exit of users.

\subsection{Outline of The Design Framework}

\begin{figure}
\centering
\includegraphics[width =4.0in]{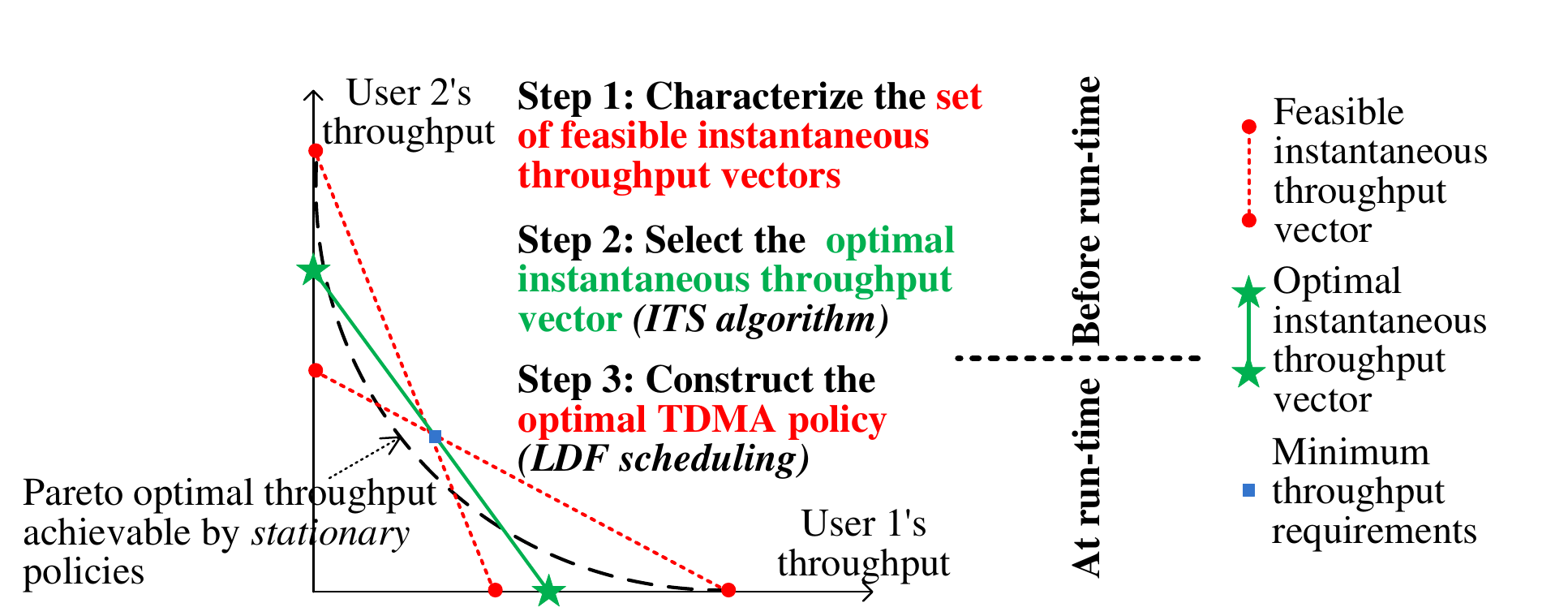}
\caption{The design framework to solve the policy design problem. The feasible instantaneous throughput vectors lie in different hyperplanes (red
dash lines) that go through the vector of minimum throughput requirements (the blue square). This results in the key difference from the design
framework in \cite[Fig.~3]{XiaoMihaela_CognitiveRadio}. In \cite{XiaoMihaela_CognitiveRadio}, all the feasible instantaneous throughput vectors lie
in one hyperplane.} \label{fig:DesignFramework_StrongNegativeExternality}
\end{figure}

The protocol design problem \eqref{eqn:PolicyDesignProblem} is difficult to solve directly, because the decision variable $\bm{\pi}$ is the spectrum
sharing policy, which is a mapping from the set of all histories to the set of actions. We first unravel an important property of the optimal TDMA
policy, namely each user should adopt the same power level whenever it transmits (see Lemma~\ref{lemma:EqualPowerLevel}). This greatly reduces the
dimension of the decision variable; now we only need to find the \emph{single} transmit power level (or equivalently, the instantaneous throughput)
of each user and the transmission schedule. We propose a three-step design framework, illustrated in
Fig.~\ref{fig:DesignFramework_StrongNegativeExternality}, to solve the design problem. First, we characterization of the set of feasible
instantaneous throughput vectors under which the users can fulfill their throughput requirements (see
Theorem~\ref{theorem:CharacterizeEquilibriumPayoff}). Based on this, we then reformulate the original problem \eqref{eqn:PolicyDesignProblem} into a
problem of finding the optimal instantaneous throughput vector, and propose a distributed instantaneous throughput selection (ITS) algorithm to solve
the reformulated problem (see Theorem~\ref{theorem:ITS}). Finally, given the optimal instantaneous throughput vector, we propose a
longest-distance-first (LDF) scheduling algorithm to determine the transmission schedule, which results in the optimal TDMA policy that solves the
design problem \eqref{eqn:PolicyDesignProblem} (see Theorem~\ref{theorem:EquilibriumStrategy}). We illustrate the design framework in
Fig.~\ref{fig:DesignFramework_StrongNegativeExternality}.

\subsection{Solving The Policy Design Problem}
We first prove a key property of the optimal energy-efficient TDMA protocol: each user should choose the same power level whenever it transmits.
\begin{lemma}\label{lemma:EqualPowerLevel}
The optimal solution $\bm{\pi}^*$ to the design problem \eqref{eqn:PolicyDesignProblem} must satisfy that each user $i$ chooses the same power level
whenever it transmits, namely $\pi_i^*(t_1)=\pi_i^*(t_2)$ for all $t_1$ and $t_2$ such that $\pi_i^*(t_1)>0$ and $\pi_i^*(t_2)>0$.
\end{lemma}
\begin{IEEEproof}
See Appendix~\ref{proof:EqualPowerLevel}.
\end{IEEEproof}

Lemma~\ref{lemma:EqualPowerLevel} greatly simplifies the design problem: now we only need to find a single optimal power level $p_i^*$ for each user
$i$ to choose whenever it transmits, instead of solving for its optimal power levels in all its transmissions. In the following, we first find the
optimal power levels $\{p_i^*\}_{i\in\mathcal{M}\cup\mathcal{N}}$ during the users' transmissions (which is equivalent to finding each user $i$'s
optimal instantaneous throughput, $r_i^*\triangleq\log_2\left(1+\frac{g_{ii}p_i^*}{\sigma_i^2}\right)$). Then given
$\{p_i^*\}_{i\in\mathcal{M}\cup\mathcal{N}}$ (or $\{r_i^*\}_{i\in\mathcal{M}\cup\mathcal{N}}$), we find the transmission schedule that achieves the
minimum throughput requirements.

\subsubsection{Step 1 -- Characterizing feasible instantaneous throughput vectors}
Now we formulate the problem of finding the users' optimal instantaneous throughput $\{r_i^*\}_{i\in\mathcal{M}\cup\mathcal{N}}$. First, the
structure of the optimal TDMA protocol discovered in Lemma~\ref{lemma:EqualPowerLevel} enables us to establish the following relationship between the
average throughput and the average energy consumption:
\begin{eqnarray}\label{eqn:Relationship_ThroughputEnergy}
&\frac{P_i(\pi_i)}{R_i(\bm{\pi})} = \frac{(1-\delta)\sum_{t=0}^\infty \delta^t \bm{1}_{\{\pi_i(t)>0\}} p_i^{\rm tdma}}{(1-\delta)\sum_{t=0}^\infty
\delta^t \bm{1}_{\{\pi_i(t)>0\}} \log_2\left(1+\frac{g_{ii}p_i^{\rm tdma}}{\sigma_i^2}\right)} = \frac{p_i^{\rm
tdma}}{\log_2\left(1+\frac{g_{ii}p_i^{\rm tdma}}{\sigma_i^2}\right)} = \frac{\sigma_i^2}{g_{ii}}\cdot\frac{2^{r_i^{\rm tdma}}-1}{r_i^{\rm tdma}},
\end{eqnarray}
where $\bm{1}_{\{\cdot\}}$ is the indicator function, $p_i^{\rm tdma}$ is user $i$'s power level when it transmits in the TDMA protocol, and
$r_i^{\rm tdma}$ is the corresponding instantaneous throughput. We can see from \eqref{eqn:Relationship_ThroughputEnergy} that given $r_i^{\rm
tdma}$, the average energy consumption $P_i(\pi_i)$ is proportional to the average throughput $R_i(\bm{\pi})$. Hence, to minimize the energy
consumption, we should let $R_i(\bm{\pi})=R_i^{\rm min}$ for all $i$. Then based on \eqref{eqn:Relationship_ThroughputEnergy}, we can rewrite the
objective function $E(P_1(\pi_1),\ldots,P_{M+N}(\pi_{M+N}))$ of the design problem \eqref{eqn:PolicyDesignProblem} as a function of the instantaneous
throughput $\{r_i^{\rm tdma}\}_{i\in\mathcal{M}\cup\mathcal{N}}$:
\begin{eqnarray}
&E\left(\frac{\sigma_1^2}{g_{11}}\cdot\frac{2^{r_1^{\rm tdma}}-1}{r_1^{\rm tdma}}\cdot R_1^{\rm
min},\ldots,\frac{\sigma_{M+N}^2}{g_{{M+N}{M+N}}}\cdot\frac{2^{r_{M+N}^{\rm tdma}}-1}{r_{M+N}^{\rm tdma}}\cdot R_{M+N}^{\rm min}\right).\nonumber
\end{eqnarray}

An instantaneous throughput vector $\{r_i^{\rm tdma}\}_{i\in\mathcal{M}\cup\mathcal{N}}$ is feasible, if there exists a TDMA protocol $\bm{\pi}$ that
has the instantaneous throughput $\{r_i^{\rm tdma}\}_{i\in\mathcal{M}\cup\mathcal{N}}$ and can achieve the minimum average throughput $\{R_i^{\rm
min}\}_{i\in\mathcal{M}\cup\mathcal{N}}$. Before characterizing the feasible instantaneous throughput vectors, we write $\bm{\tilde{p}}^i=(p_i^{\rm
tdma}(r_i^{\rm tdma}), \bm{p}_{-i}=\bm{0})$ as the joint power profile when user $i$ transmits in a TDMA policy. Now we state
Theorem~\ref{theorem:CharacterizeEquilibriumPayoff}.
\begin{theorem}\label{theorem:CharacterizeEquilibriumPayoff}
An instantaneous throughput vector $\{r_i^{\rm tdma}\}_{i\in\mathcal{M}\cup\mathcal{N}}$ is feasible for the minimum throughput requirements
$\{R_i^{\rm min}\}_{i\in\mathcal{M}\cup\mathcal{N}}$, if the following conditions are satisfied:
\begin{itemize}
\item Condition~1: the discount factor $\delta$ satisfies $\delta \geq \underline{\delta} \triangleq 1/\left(1+\frac{1-\sum_{i\in\mathcal{M}\cup\mathcal{N}}
\underline{\mu}_i}{M+N-1+\sum_{i\in\mathcal{M}\cup\mathcal{N}}\sum_{j\neq i} (-\rho(y=1|\bm{\tilde{p}}^i)/b_{ij})}\right)$,
where $b_{ij} = \sup_{p_j\in \mathcal{P}_j, p_j\neq \tilde{p}_j^i}
\frac{\rho(y=1|\bm{\tilde{p}}^i)-\rho(y=1|p_j,\bm{\tilde{p}}_{-j}^i)}{r_j(p_j,\bm{\tilde{p}}_{-j}^i)/\bar{r}_j}$, and $\underline{\mu}_i \triangleq
\max_{j\neq i} \frac{1-\rho(y=1|\bm{\tilde{p}}^i)}{-b_{ij}}$.
\item Condition~2: $\sum_{i\in\mathcal{M}\cup\mathcal{N}} R_i^{\rm min}/r_i^{\rm tdma} = 1$, and $r_i^{\rm tdma} \leq R_i^{\rm min}/\underline{\mu}_i$.
\end{itemize}
\end{theorem}
\begin{IEEEproof}
See Appendix~\ref{proof:CharacterizeEquilibriumPayoff}.
\end{IEEEproof}

The problem of finding the optimal instantaneous throughput can then be formulated as
\begin{eqnarray}\label{eqn:PolicyDesignProblem_OptimalThroughput}
\{r_i^*\}_{i\in\mathcal{M}\cup\mathcal{N}}=\!\!\!\!&\displaystyle\arg\min_{\{r_i^{\rm tdma}\}_{i\in\mathcal{M}\cup\mathcal{N}}}&
\!\!\!\!\!\!E\!\left(\left\{\frac{\sigma_i^2}{g_{ii}} \frac{2^{r_i^{\rm tdma}}-1}{r_i^{\rm tdma}} R_i^{\rm
min}\right\}_{i\in\mathcal{M}\cup\mathcal{N}}\right) \nonumber\\
&s.t.& \!\!\!\!\!\!\sum_{i\in\mathcal{M}\cup\mathcal{N}} \frac{R_i^{\rm min}}{r_i^{\rm tdma}}=1, \\
&    & \!\!\!\!\!\!0< r_i^{\rm tdma}\leq \bar{r}_i \triangleq R_i^{\rm min}/\underline{\mu}_i,~\forall i\in\mathcal{M}\cup\mathcal{N}. \nonumber
\end{eqnarray}

\subsubsection{Step 2 -- Select the optimal instantaneous throughput vector}
We solve the above optimization problem \eqref{eqn:PolicyDesignProblem_OptimalThroughput} for the optimal instantaneous throughput vector
$\{r_i^*\}_{i\in\mathcal{M}\cup\mathcal{N}}$ using the distributed ITS algorithm, which is proved to converge in logarithmic time in
Theorem~\ref{theorem:ITS}.

The ITS algorithm essentially solves the following equation (derived from the KKT condition) in a distributed fashion:
\begin{eqnarray}\label{eqn:EquationToSolveInITS}
\frac{\partial E}{\partial P_i}|_{P_i=\frac{\sigma_i^2 R_i^{\rm min}}{g_{ii}} \frac{2^{r_i^*}-1}{r_i^*}} \cdot \left(2^{r_i^*}-1- \ln2 \cdot r_i^*
\cdot 2^{r_i^*}\right) \cdot \frac{\sigma_i^2}{g_{ii}}= -\lambda,
\end{eqnarray}
where $\lambda$ is the Lagrangian multiplier for the constraint $\sum_{i} \frac{R_i^{\rm min}}{r_i^{\rm tdma}}=1$ in
\eqref{eqn:PolicyDesignProblem_OptimalThroughput}, and should be chosen such that $\sum_{i} \frac{R_i^{\rm min}}{r_i^*}=1$. The term $\frac{\partial
E}{\partial P_i}$ in \eqref{eqn:EquationToSolveInITS} is the derivative of the energy efficiency criterion $E(\cdot)$ with respect to user $i$'s
average energy consumption. If the energy efficiency criterion is the weighted sum of all the users' energy consumptions, we have $\frac{\partial
E}{\partial P_i}|_{P_i=\frac{\sigma_i^2 R_i^{\rm min}}{g_{ii}} \frac{2^{r_i^*}-1}{r_i^*}} = w_i,~\forall r_i^*$. If the energy efficiency criterion
is the weighted proportional fairness $-\sum_{i\in\mathcal{M}\cup\mathcal{N}} w_i \log(P_i)$, we have $\frac{\partial E}{\partial
P_i}|_{P_i=\frac{\sigma_i^2 R_i^{\rm min}}{g_{ii}} \frac{2^{r_i^*}-1}{r_i^*}} = -w_i \frac{g_{ii}}{\sigma_i^2 R_i^{\rm min}} \cdot
\frac{r_i^*}{2^{r_i^*}-1}$. Each user $i$ selects the term $\frac{\partial E}{\partial P_i}$ in the ITS algorithm based on the energy efficiency
criterion chosen by the protocol designer.


\begin{algorithm}
\caption{Instantaneous Throughput Selection (ITS) algorithm run by user $i$.} \label{table:ITS}
\begin{algorithmic}[1]
\REQUIRE Minimum throughput requirement $R_i^{\rm min}$, precision $e$

\STATE Set $\underline{\lambda}=0$, $\bar{\lambda}=1$, $\lambda=\bar{\lambda}$.

\STATE Solve \eqref{eqn:EquationToSolveInITS} for $r_i^*$, set $r_i^*\leftarrow\min\{r_i^*,\bar{r}_i\}$

\STATE Broadcast $R_i^{\rm min}/r_i^*$, and receive $R_j^{\rm min}/r_j^*$ for all $j\neq i$

\WHILE{$\sum_{j\in\mathcal{M}\cup\mathcal{N}} R_j^{\rm min}/r_j^* > 1$} 

\STATE $\bar{\lambda}\leftarrow 2\cdot\bar{\lambda}$, $\lambda\leftarrow\bar{\lambda}$

\STATE Solve \eqref{eqn:EquationToSolveInITS} for $r_i^*$, set $r_i^*\leftarrow\min\{r_i^*,\bar{r}_i\}$

\STATE Broadcast $R_i^{\rm min}/r_i^*$, and receive $R_j^{\rm min}/r_j^*$ for all $j\neq i$

\ENDWHILE

\WHILE{$\left|\sum_{j\in\mathcal{M}\cup\mathcal{N}} R_j^{\rm min}/r_j^* - 1\right|>e$} 

\STATE $\lambda\leftarrow\frac{\underline{\lambda}+\bar{\lambda}}{2}$

\STATE Solve \eqref{eqn:EquationToSolveInITS} for $r_i^*$, set $r_i^*\leftarrow\min\{r_i^*,\bar{r}_i\}$

\STATE Broadcast $R_i^{\rm min}/r_i^*$, and receive $R_j^{\rm min}/r_j^*$ for all $j\neq i$

\IF{$\sum_{j\in\mathcal{M}\cup\mathcal{N}} R_j^{\rm min}/r_j^* < 1$}

\STATE $\bar{\lambda}\leftarrow\lambda$

\ELSE

\STATE $\underline{\lambda}\leftarrow\lambda$

\ENDIF

\ENDWHILE

\STATE Normalize $r_i^*\leftarrow r_i^*/\left(\sum_{j\in\mathcal{M}\cup\mathcal{N}} R_j^{\rm min}/r_j^*\right)$

\end{algorithmic}
\end{algorithm}

\begin{theorem}\label{theorem:ITS}
The problem \eqref{eqn:PolicyDesignProblem_OptimalThroughput} of finding the optimal instantaneous throughput vector can be converted into a convex
optimization problem, whose solution $\{r_i^*\}_{i\in\mathcal{M}\cup\mathcal{N}}$ can be found by each user running the distributed ITS algorithm.
The algorithm converges linearly\footnote{Following \cite[Sec.~9.3.1]{Boyd}, we define linear convergence as follows. Suppose that the sequence
$\{x_k\}$ converges to $x$. We say that this sequence converges linearly at rate $c$, if we have $\lim_{k\rightarrow\infty}
\frac{|x_{k+1}-x|}{|x_k-x|} = c$.} at rate $\frac{1}{2}$.
\end{theorem}
\begin{IEEEproof}
See Appendix~\ref{proof:ITS}.
\end{IEEEproof}

\begin{algorithm}
\caption{The Longest-Distance-First (LDF) scheduling run by user $i$.} \label{table:EquilibriumStrategy}
\begin{algorithmic}
\REQUIRE $\{R_j^{\rm min}/r_j^\star\}_{j\in\mathcal{M}\cup\mathcal{N}}$, $r_i^*$

\STATE \textbf{Initialization:} Set $t=0$, $r_j^\prime(0)=R_j^{\rm min}/r_j^*$ for all $j\in\mathcal{M}\cup\mathcal{N}$

\REPEAT

\STATE Calculates the distance from the optimal operating point $d_j(t) =
\frac{r_j^\prime(t)-\underline{\mu}_j}{1-r_j^\prime(t)}\rho(y=1|\mathbf{\tilde{p}}^j),\forall j$

\STATE Find the user with the largest distance $i^*\triangleq\arg\max_{j\in\mathcal{M}\cup\mathcal{N}} d_j(t)$

\IF{$i=i^*$}

\STATE Transmit at power level $p_i^{\rm tdma}(r_i^*)$ \\

\ENDIF

\STATE Updates $r_j^\prime(t+1)$ for all $j\in\mathcal{M}\cup\mathcal{N}$ as follows:

\IF{No Distress Signal Received At Time Slot $t$}

\STATE $r_{i^*}^\prime(t+1)=\frac{1}{\delta}\cdot r_{i^*}^\prime(t)-(\frac{1}{\delta}-1)\cdot(1+\sum_{j\neq {i^*}}
\frac{\rho(y=1|\mathbf{\tilde{p}}^{i^*})}{-b_{{i^*}j}})$

\STATE $r_j^\prime(t+1)=\frac{1}{\delta}\cdot r_j^\prime(t)+(\frac{1}{\delta}-1)\cdot\frac{\rho(y=1|\mathbf{\tilde{p}}^{i^*})}{-b_{i^* j}},\forall
j\neq i^*$

\ELSE

\STATE $r_{i^*}^\prime(t+1)=\frac{1}{\delta}\cdot r_{i^*}^\prime(t)-(\frac{1}{\delta}-1)\cdot(1-\sum_{j\neq {i^*}}
\frac{\rho(y=0|\mathbf{\tilde{p}}^{i^*})}{-b_{{i^*}j}})$

\STATE $r_j^\prime(t+1)=\frac{1}{\delta}\cdot r_j^\prime(t)-(\frac{1}{\delta}-1)\cdot\frac{\rho(y=0|\mathbf{\tilde{p}}^{i^*})}{-b_{i^* j}},\forall
j\neq i^*$

\ENDIF

\STATE $t\leftarrow t+1$

\UNTIL{$\varnothing$}
\end{algorithmic}
\end{algorithm}

\subsubsection{Step 3 -- Construct the optimal deviation-proof policy} Given the optimal instantaneous throughput vector, each user $i$ runs the
longest-distance-first scheduling algorithm in a decentralized manner. On one hand, the transmission schedule can be viewed as a simple
``largest-distance-first'' scheduling, namely the user farthest away from its throughput requirement transmits. On the other hand, it is nontrivial
to define the ``distance'' from its throughput requirement. As we will prove later, user $j$'s distance from its throughput requirement can be
defined as $d_j(t) = \frac{r_j^\prime(t)-\underline{\mu}_j}{1-r_j^\prime(t)+\sum_{k\neq j} (-\rho(y=1|\bm{\tilde{p}}^j)/b_{jk})}$, where
$r_j^\prime(t)$ is the future throughput to achieve starting from time slot $t$ normalized by $r_j^*$. The normalized future throughput
$r_j^\prime(t)$ can be also interpreted the future transmission opportunity. If user $j$ transmitted all the time in the future, it would have an
average throughput $r_j^*$. If it transmits in a fraction $r_j^\prime(t)$ of time after time $t$, it has an average future throughput of
$r_j^\prime(t) \cdot r_j^*$.

Theorem~\ref{theorem:EquilibriumStrategy} proves the desirable properties of the LDF scheduling algorithm.
\begin{theorem}\label{theorem:EquilibriumStrategy}
If each user $i\in\mathcal{M}\cup\mathcal{N}$ runs the LDF scheduling algorithm, then we have
\begin{itemize}
\item each user $i$ can achieve its minimum throughput requirement $R_i^{\rm min}$ with an energy consumption $P_i$ that minimizes the energy
efficiency criterion $E(P_1, \ldots, P_{M+N})$;
\item if a user does not
follow the algorithm, it will either fail to achieve the minimum throughput requirement, or achieve it with a higher energy consumption;
\item the distance between each user $i$'s average throughput at time $t$ and its throughput requirement decreases exponentially with time, namely
\begin{eqnarray}
|(1-\delta) \sum_{\tau=0}^t \delta^\tau \cdot r_i^\tau - R_i^{\rm min}| \leq r_i^*\cdot\delta^{t+1}.
\end{eqnarray}
\end{itemize}
\end{theorem}
\begin{IEEEproof}
See Appendix~\ref{proof:EquilibriumStrategy}.
\end{IEEEproof}

Theorems~\ref{theorem:ITS} and \ref{theorem:EquilibriumStrategy} establish the convergence results of our proposed scheme. Theorem~\ref{theorem:ITS}
proves that the process of finding the optimal instantaneous throughput vector converges in logarithmic time, and
Theorem~\ref{theorem:EquilibriumStrategy} proves that the LDF scheduling achieves the minimum throughput requirements in logarithmic time. Hence, the
overall convergence speed is fast.

Note that our convergence results are very different from the convergence results in some recent works on power control in cognitive radio
\cite{ZhengTan_JSAC2013} and wireless networks \cite{TanChiangSrikant_TON2013}. These works \cite{ZhengTan_JSAC2013}\cite{TanChiangSrikant_TON2013}
belong to the stationary spectrum sharing policies, namely they aim to find the optimal fixed power levels of the users that maximize the network
utility. The convergence results in \cite{ZhengTan_JSAC2013}\cite{TanChiangSrikant_TON2013} differ from our results in two important ways. First,
since our work studies nonstationary spectrum sharing with time-varying power levels, we need to determine not only the optimal power levels of the
users, but also the transmission schedule of the users. We prove that the average throughput obtained by adopting the proposed LDF scheduling
converges linearly. Such a result does not appear in \cite{ZhengTan_JSAC2013}\cite{TanChiangSrikant_TON2013}. Second, the techniques used in proving
the convergence to the optimal power levels are different. In \cite{ZhengTan_JSAC2013}\cite{TanChiangSrikant_TON2013}, the algorithms are akin to the
celebrated distributed power control algorithm \cite{Yates95}, and hence the proofs use and extend the ``standard interference function'' argument.
Such an argument is not used in our work since there is no interference among the users under the proposed TDMA spectrum sharing policy.


\subsection{Implementation}
We discuss the total overhead of information exchange and feedback and the computational complexity of the proposed scheme.


\begin{table}\scriptsize
\renewcommand{\arraystretch}{0.9}
\caption{Comparison of the total overhead of initial information exchange and feedback.} \label{table:InformationExchangeOverhead} \centering
\begin{tabular}{|p{1.0cm}|p{7cm}|p{7cm}|}
\hline
 & Information exchange before run-time & Feedback at run-time \\
\hline
\multirow{2}{*}{\cite{Yates95}--\cite{TanPalomarChiang_TON2009}} & \multirow{2}{*}{N/A} & Each user $i$: $I_{-i}$ \emph{each time slot} \\
                                                                     &  & Amount: $M+N$ real numbers in \emph{each time slot} \\
\hline
\multirow{2}{*}{\cite{SorooshyariTanChiang}} & A spectrum coordinator to each user $i$: degradation of its minimum throughput requirement & Each user $i$: $I_{-i}$ in \emph{each time slot}, each PU: distress signal when necessary \\
                                             & Amount: $M+N$ real numbers                                         & Amount: $M+N$ real numbers in \emph{each time slot}, a distress signal when necessary \\
\hline
\multirow{2}{*}{Proposed} & Each user $i$ broadcasts to all the other users: $\rho(y=1|\mathbf{\tilde{p}}^i)$ and $\{b_{ji}\}_{j\neq i}$ once, and $R_i^{\rm min}/r_i^*$ at each iteration of the ITS algorithm; Each user $i$'s receiver to its transmitter: $\bar{I}_i,\underline{I}_i$ & distress signal when necessary \\
                          & Amount: $(M+N)^2+(M+N)\cdot \mathcal{O}(\log_2(1/e))$ real numbers                                                                & Amount: a distress signal when necessary \\
\hline
\end{tabular}
\end{table}

\subsubsection{Overhead of initial information exchange and feedback}
In Table~\ref{table:InformationExchangeOverhead}, we compare the overhead of information exchange and feedback of the proposed framework with the
energy efficient spectrum sharing policies proposed in \cite{Yates95}--\cite{TanPalomarChiang_TON2009} and \cite{SorooshyariTanChiang} for wireless
networks and cognitive radio networks, respectively. Before run-time, the information exchange in the proposed framework comes from the ITS algorithm
($(M+N)\cdot \mathcal{O}(\log_2(1/e))$ with $e$ being the performance loss tolerance) and the exchange of $b_{ij}$ for the LDF scheduling. The
exchange of $b_{ij}$ is for deviation-proofness. However, in the run time, the feedback overhead of the proposed policy is significantly lower than
that of \cite{Yates95}--\cite{SorooshyariTanChiang}. Specifically, in \cite{Yates95}--\cite{SorooshyariTanChiang}, each user $i$'s receiver needs to
feedback the interference temperature $I_{-i}$ in \emph{each time slot}. Hence, the total amount of feedback in
\cite{Yates95}--\cite{SorooshyariTanChiang} grows linearly with time. In conclusion, our proposed framework has a much lower total overhead than
\cite{Yates95}--\cite{SorooshyariTanChiang}.

\subsubsection{Computational complexity} The implementation of the proposed policy includes the ITS algorithm before run-time and the LDF scheduling
at run-time. First, both the ITS algorithm and the LDF scheduling converge fast in logarithmic time as proved in Theorems~\ref{theorem:ITS} and
\ref{theorem:EquilibriumStrategy}. Second, each iteration in the ITS algorithm involves solving the equation \eqref{eqn:EquationToSolveInITS}, which
can be done efficiently using the Newton method. Each iteration in the LDF scheduling involves computing $M+N$ indices
$\{d_j(t)\}_{j\in\mathcal{M}\cup\mathcal{N}}$ and $M+N$ normalized values $\{r_j^\prime(t)\}_{j\in\mathcal{M}\cup\mathcal{N}}$, all of which are
determined by analytical expressions. Finally, although the original definition of the policy requires each user to memorize the entire history of
distress signals, in the LDF scheduling, each user only needs to know the current distress signal $y^t$ and memorize $M+N$ normalized values
$\{r_j^\prime(t)\}_{j\in\mathcal{M}\cup\mathcal{N}}$. In conclusion, the overall computational complexity of each user in implementing the proposed
policy is small.

\subsection{Users Entering and Leaving the Network}
\begin{figure} \centering
\includegraphics[width =4.0in]{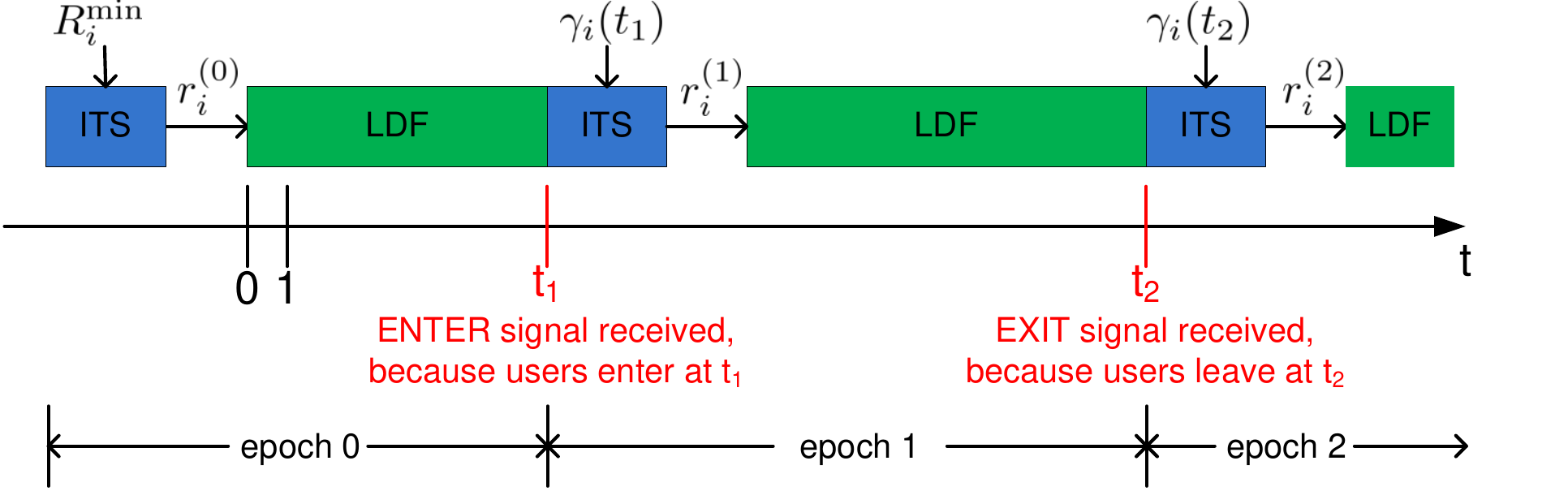}
\caption{The proposed protocol implemented by user $i$ when it receives ``ENTER'' signal at $t_1$ and ``EXIT'' signal at $t_2$.}
\label{fig:PowerControlProtocol_UserEnterLeave}
\end{figure}

We adapt the protocol to the scenario where users enter and leave the network. We divide time into \emph{epochs}, where a new epoch begins when users
enter or leave. The system starts at epoch 0, and we denote the optimal instantaneous throughput in epoch 0 by $r_i^{(0)}$. When new users enter or
existing users leave at $t_1$, each of them broadcasts a ``ENTER'' or ``EXIT'' signal, respectively. Upon receiving such a signal, the users run the
ITS algorithm again to determine the optimal instantaneous throughput in epoch 1, $r_i^{(1)}$. Note that for each existing user $i$, the input to the
ITS algorithm is the continuation throughput at $t_1$, namely $\gamma_i(t_1)$; while for each new user $j$, the input should be its minimum
throughput $R_j^{\rm min}$. Then they run the LDF scheduling with the new instantaneous throughput, until a new epoch begins when the ``ENTER'' or
``EXIT'' signals are broadcast by some users at $t_2$. We illustrate how to adapt the protocol in Fig.~\ref{fig:PowerControlProtocol_UserEnterLeave}.

One nice property of the proposed protocol is that, the convergence of the LDF scheduling is not affected by users coming or leaving.

\begin{theorem}\label{theorem:DynamicEntryExit}
In the proposed spectrum sharing protocol, each user's average throughput converges to the minimum throughput requirement in logarithmic time, even
with users entering and leaving the network.
\end{theorem}
\begin{IEEEproof}
See Appendix~\ref{proof:DynamicEntryExit}.
\end{IEEEproof}

Note that we can also deal with the changes of system parameters (e.g. the channel gains) in the same way as we deal with the dynamic entry and exit
of users. Specifically, whenever a user observes a change in the system parameters, it can broadcast a signal that triggers the users to run the ITS
algorithm and the LDF scheduling again. The convergence result in Theorem~\ref{theorem:DynamicEntryExit} also applies to this case.

In some works \cite{TanPalomarChiang_TON2009} for energy efficient power control in wireless networks, the locally stable asymptotic convergence of
the proposed algorithm is proved. The locally stable asymptotic convergence guarantees that slight perturbation from the equilibrium (induced by, for
example, an incoming user) will not make the algorithm diverge. However, the convergence result in Theorem~\ref{theorem:DynamicEntryExit} are
different from that in \cite{TanPalomarChiang_TON2009}. Specifically, we study the convergence of not only the transmit power levels, but also the
transmission schedule, which is not studied in \cite{TanPalomarChiang_TON2009}. More importantly, the influence of dynamic entry and exit of users on
the convergence and stability is quite different in our work as compared to \cite{TanPalomarChiang_TON2009}. Since our proposed policy is TDMA, there
is no interference among the users. Hence, an incoming user will not interfere with the existing users when they transmit. In other words, the
influence of incoming users is not through the interference as in \cite{TanPalomarChiang_TON2009}, but through acquiring the transmission
opportunities of the existing users. We show that under such perturbation (in terms of transmission opportunities), the proposed LDF scheduling still
converges to the target throughput at the same rate.

\section{Performance Evaluation}\label{sec:Simulation}
In this section, we demonstrate the performance gain of our spectrum sharing policy over existing policies, and validate our theoretical analysis
through numerical results. Throughout this section, we use the following system parameters by default unless we change some of them explicitly. The
noise powers at all the users' receivers are $0.05$~W. For simplicity, we assume that the direct channel gains have the same distribution
$g_{ii}\thicksim\mathcal{CN}(0,1), \forall i$, and the cross channel gains have the same distribution $g_{ij}\thicksim\mathcal{CN}(0,\alpha), \forall
i\neq j$, where $\alpha$ is defined as the \emph{cross interference level}. The interference temperature threshold is $I=1$~W. The measurement error
$\varepsilon$ is Gaussian distributed with zeros mean and variance $0.1$. The energy efficiency criterion is the average transmit power of each user.
The discount factor is $0.95$.

\subsection{Comparisons Against Existing Policies}
First, assuming that the population is fixed, we compare the proposed policy against the optimal stationary policy in
\cite{Yates95}--\cite{SorooshyariTanChiang} and two (adapted) versions of the punish-forgive (PF) policies in
\cite{EtkinTse}--\cite{XiaoMihaela_RepeatedGame}. Since the PF policies in \cite{EtkinTse}--\cite{XiaoMihaela_RepeatedGame} were originally proposed
for network utility maximization problems (e.g. maximizing the sum throughput), we need to adapt them to solve the energy efficiency problem in
\eqref{eqn:PolicyDesignProblem}. We describe the state-of-the-art policies that we compare against as follows.
\begin{itemize}
\item The optimal stationary policy \cite{Yates95}--\cite{SorooshyariTanChiang}: each user transmits at a fixed power level that is just large enough to fulfill the throughput requirement
under the interference from other users.
\item The (adapted) stationary punish-forgive (SPF) policy \cite{EtkinTse}--\cite{LeTreustLasaulce}: the SPF policies are dynamic policies that have two phases.
When the users have not received the distress signal, they transmit at optimal \emph{stationary} power levels. When they receive a distress signal
that indicates deviation, they switch to the punishment phase, in which all the users transmit at the Nash equilibrium power levels. In the energy
efficiency formulation, the optimal stationary power levels are the Nash equilibrium power levels. Hence, the adapted SPF policy is essentially the
same as the optimal stationary policy.
\item The adapted nonstationary punish-forgive (NPF) policy: the punish-forgive policy in \cite{XiaoMihaela_RepeatedGame} is different from those in
\cite{EtkinTse}--\cite{LeTreustLasaulce}, in that \emph{nonstationary} power levels are used when the users have not received the distress signal. In
the simulation, we adapt the NPF policy in \cite{XiaoMihaela_RepeatedGame} such that the users transmit in the same way as in the proposed policy
when they have not received the distress signal. After receiving the distress signal, the NPF policy requires the users to transmit at the optimal
stationary power levels.
\end{itemize}
Since the SPF policy is the same as the optimal stationary policy, in the rest of this section, we focus on the NPF policy, and simply refer to the
NPF policy as the PF policy.

\subsubsection{Illustrations of Different Policies}
\begin{figure}
\centering
\includegraphics[width =3.2in]{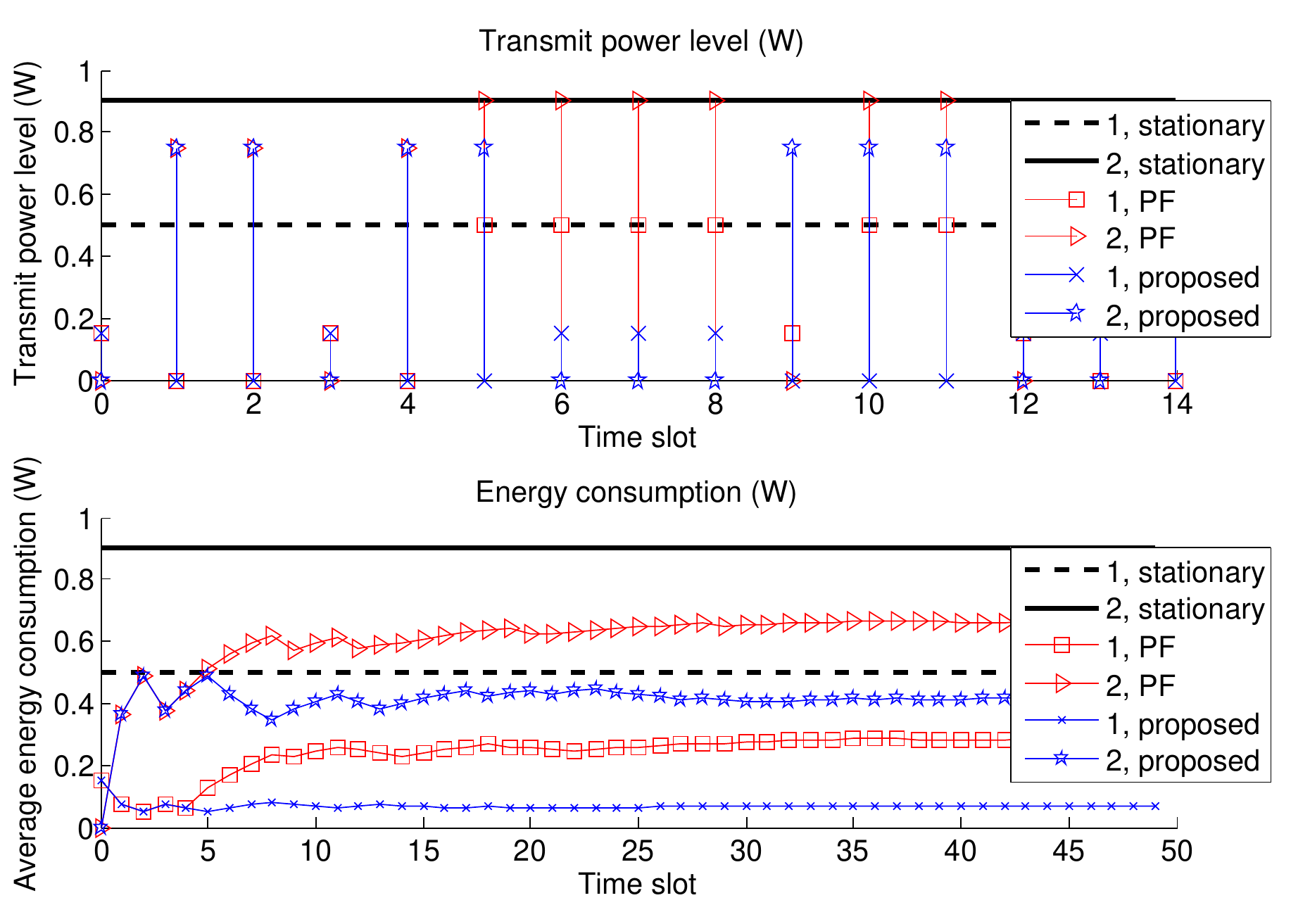}
\caption{Illustration of different policies.} \label{fig:IllustrationPolicies}
\end{figure}

Fig.~\ref{fig:IllustrationPolicies} illustrates the differences among stationary, PF, and the proposed policies in a simple case of two users, whose
minimum throughput requirements are 1~bits/s/Hz and 2~bits/s/Hz, respectively. In stationary policies, users transmit simultaneously with fixed power
levels (0.5~W and 0.9~W), which are higher than those (0.15~W and 0.75~W) in the proposed policy, because users need to overcome multi-user
interference to achieve the minimum throughput requirements. In addition, users transmit all the time in stationary polices, which results in even
higher average energy consumption.

The key difference between the proposed policy and the PF policy lies in time slot $5$, after a distress signal is sent at $t=4$. In the PF policy,
users transmit together at the same high power levels as in the stationary policy at $t=5$. In the proposed policy, user 2, the user who transmitted
at $t=4$, transmits again at $t=5$. In summary, the punishment in the PF policy is the multi-user interference, which increases the energy
consumptions of both users, while the punishment in the proposed policy is the delay in transmission, which keeps the energy consumptions low. This
advantage of the proposed policy in terms of energy efficiency is also illustrated in Fig.~\ref{fig:IllustrationPolicies}.

Finally, we can see that in the steady state, the energy consumption of the proposed policy is much lower than those in the other policies.

\subsubsection{Performance Gains}
\begin{figure}
\centering \subfloat[][Different cross interference levels.]{
\includegraphics[width =2.1in]{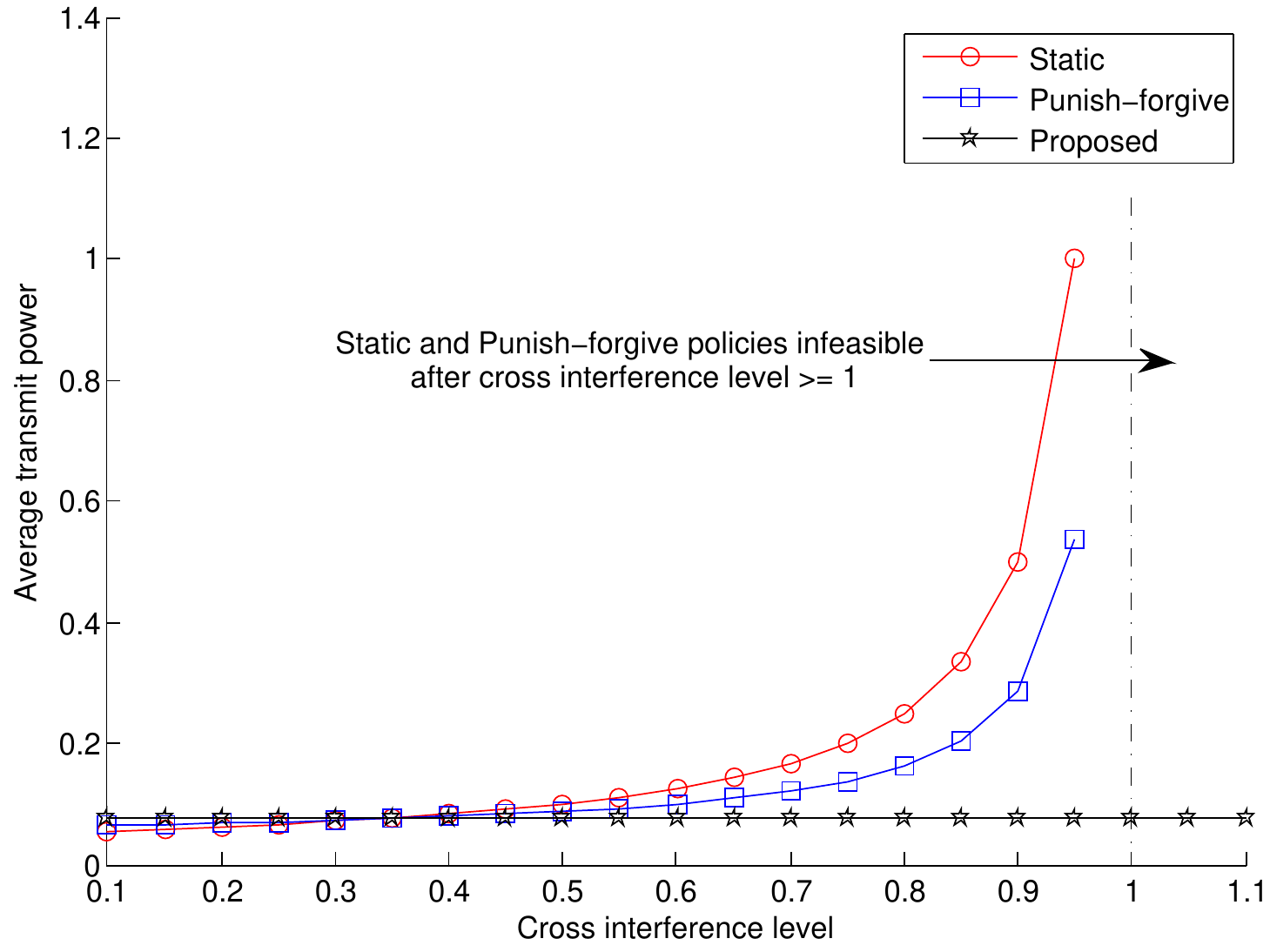}
\label{subfig:Comparison_CrossInterferenceLevel} } \subfloat[][Different numbers of users.]{
\includegraphics[width =2.1in]{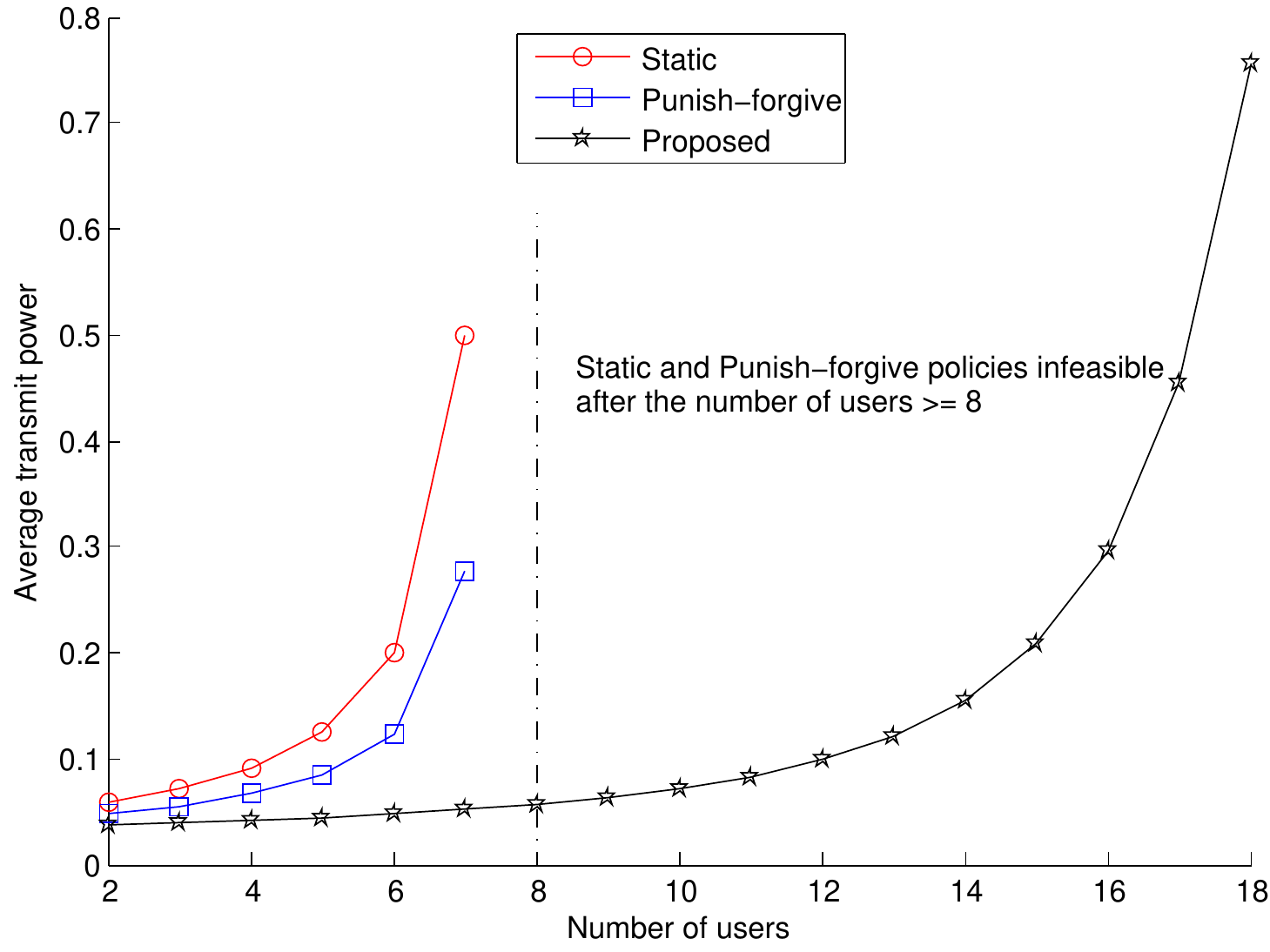}
\label{subfig:Comparison_UserNumber} } \subfloat[][Different min. throughput requirements.]{
\includegraphics[width =2.1in]{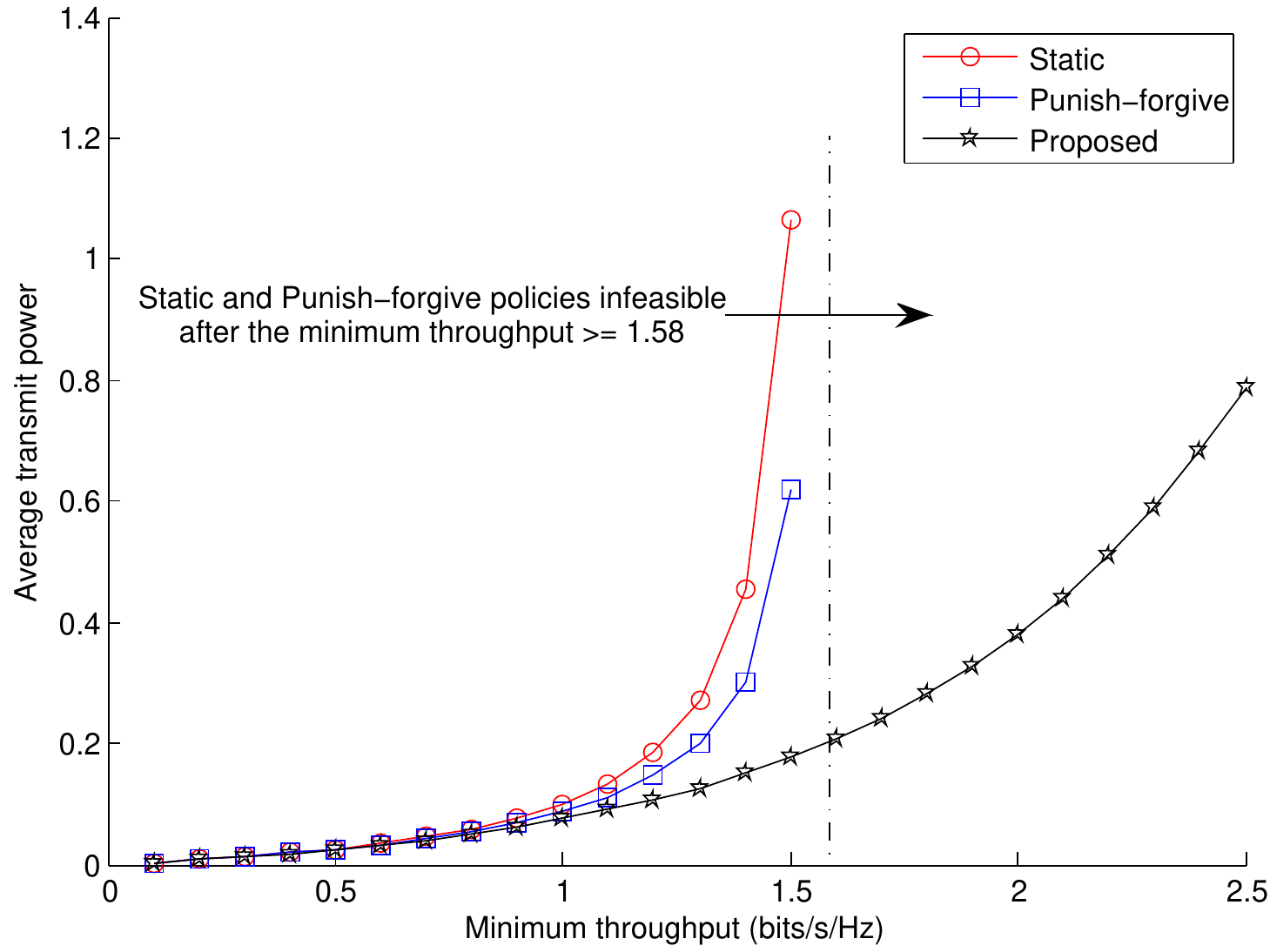}
\label{subfig:Comparison_MinimumThroughput} } \caption[]{Energy efficiency of the stationary, punish-forgive, and proposed policies under different
system parameters.}\label{fig:Comparison}
\end{figure}

We compare the energy efficiency of the optimal stationary policy, the optimal punish-forgive policy, and the proposed policy under different cross
interference levels in Fig.~\ref{subfig:Comparison_CrossInterferenceLevel}. We consider a network of two users whose minimum throughput requirements
are 1 bits/s/Hz. First, notice that the energy efficiency of the proposed policy remains constant under different cross interference levels, while
the average transmit power increases with the cross interference level in the other two policies. The proposed policy outperforms the other two
policies in medium to high cross interference levels (approximately when $\alpha\geq 0.3$). In the cases of high cross interference levels
($\alpha\geq1$), there is no stationary policy that can fulfill the minimum throughput requirements. As a consequence, the punish-forgive policies
cannot fulfill the throughput requirements when $\alpha\geq1$, either.

In Fig.~\ref{subfig:Comparison_UserNumber}, we examine how the performance of these three policies scales with the number of users. The number of
users in the network increases, while the minimum throughput requirement for each user remains 1 bits/s/Hz. The cross interference level is $\alpha =
0.2$. We can see that the stationary and punish-forgive policies are infeasible when there are more than 6 users. In contrast, the proposed policy
can accommodate 18 users in the network with each users transmitting at a power level less than 0.8~W.

Fig.~\ref{subfig:Comparison_MinimumThroughput} shows the joint spectrum and energy efficiency of the three policies. We can see that the optimal
stationary and punish-forgive polices are infeasible when the minimum throughput requirement is larger than 1.6 bits/s/Hz. On the other hand, the
proposed policy can achieve a much higher spectrum efficiency (2.5~bits/s/Hz) with a better energy efficiency (0.8~W transmit power). Under the same
average transmit power, the proposed policy is always more energy efficient than the other two policies.

In summary, the proposed policy significantly improves the spectrum and energy efficiency of existing policies in most scenarios. In particular, the
proposed policy achieves an energy saving of up to 90\%, when the cross interference level is large or the number of users is large (e.g., when
$\alpha=0.9$ in Fig.~\ref{subfig:Comparison_CrossInterferenceLevel} and when $N=7$ in Fig.~\ref{subfig:Comparison_UserNumber}). These are exactly the
deployment scenarios where improvements in spectrum and energy efficiency are much needed. In addition, the proposed policy can always remain
feasible even when the other policies cannot maintain the minimum throughput requirements.

\subsection{Adapting to Users Entering and Leaving the Network}
\begin{figure}
\centering \subfloat[][Dynamics of average energy consumption.]{
\includegraphics[width =3.2in]{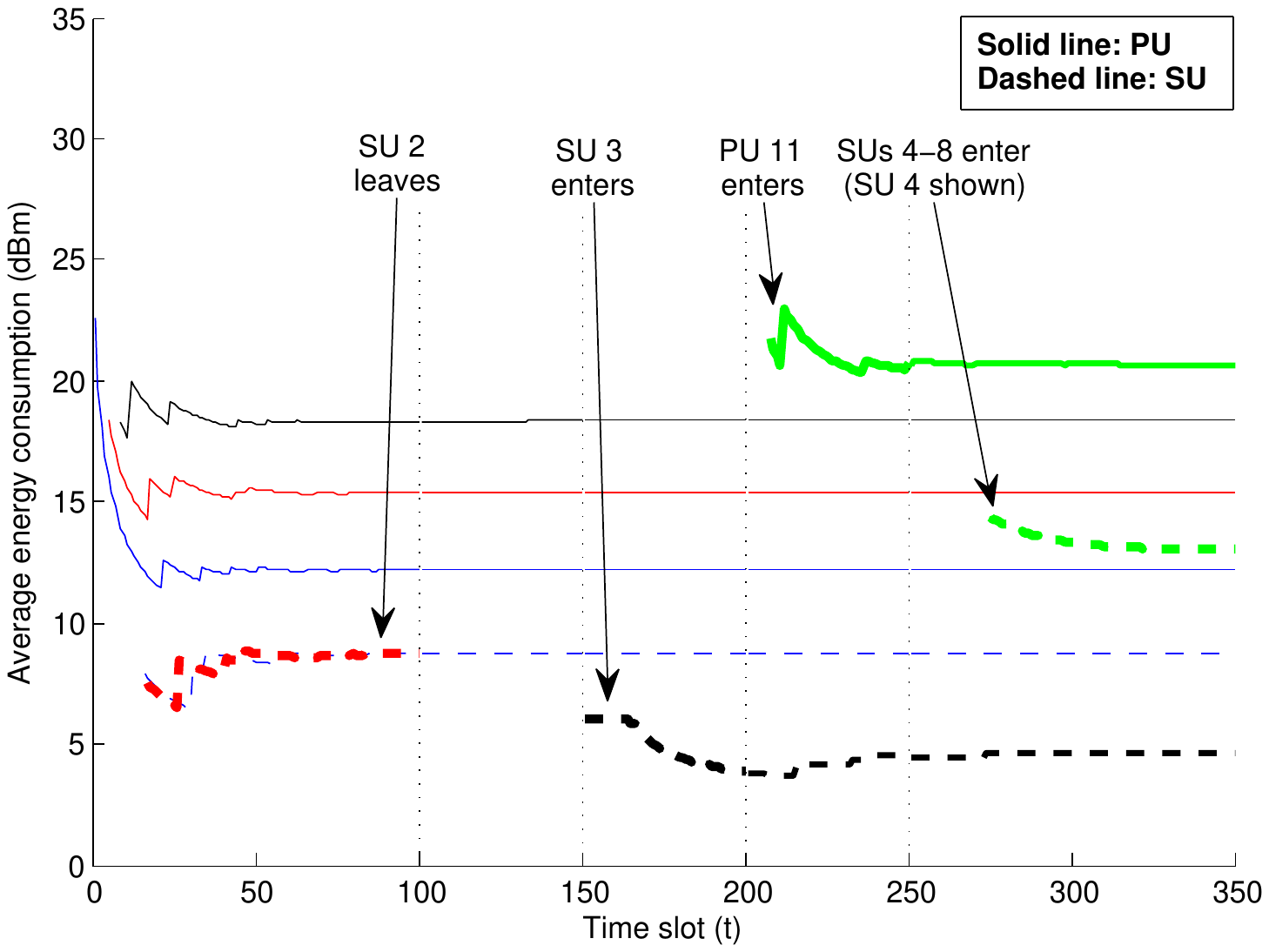}
} \subfloat[][Dynamics of average throughput.]{
\includegraphics[width =3.2in]{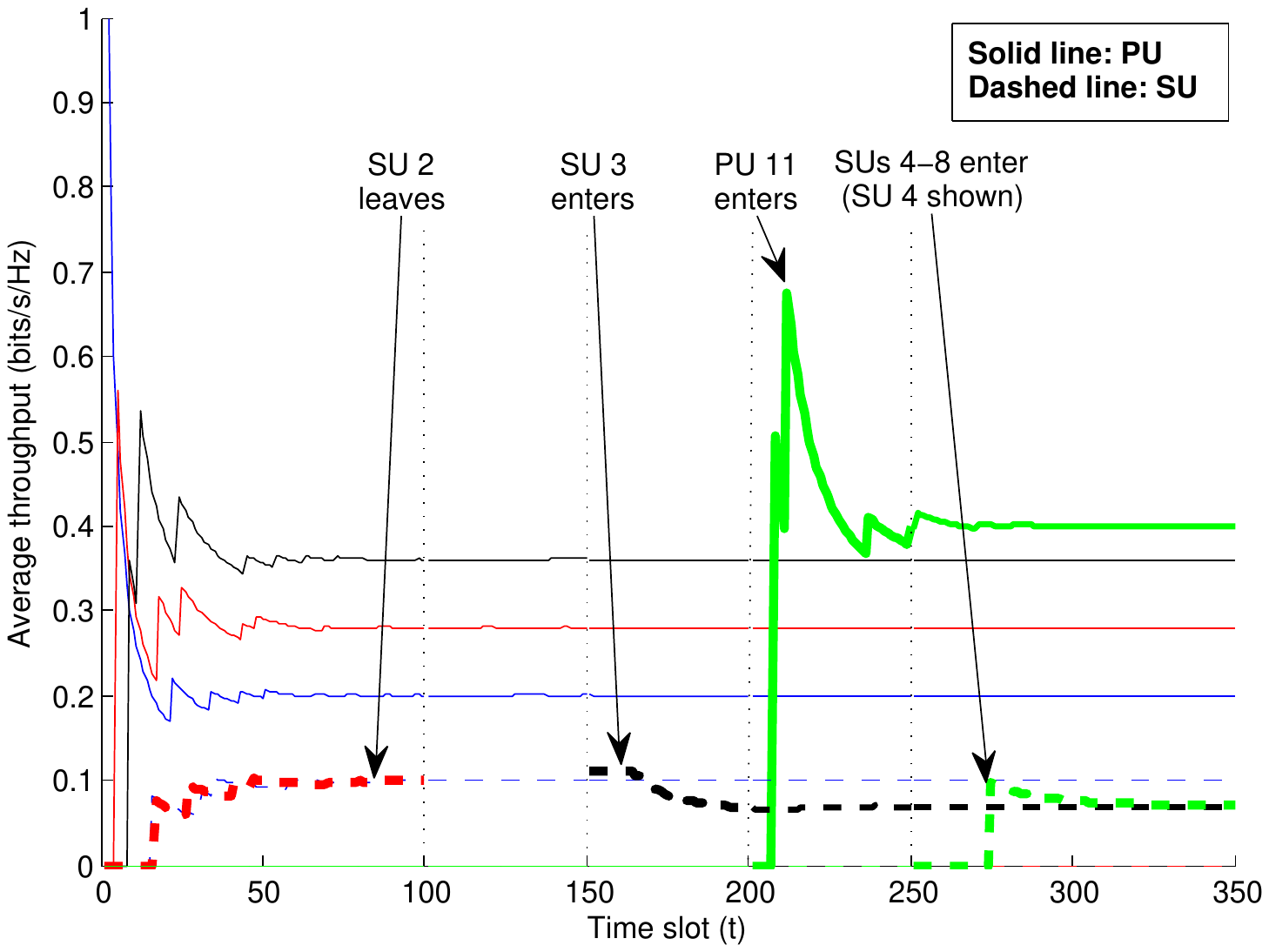}
} \caption[]{Dynamics of average energy consumption and average throughput with users entering and leaving the network. At $t=0$, there are 10 PUs
and 2 SUs. SU 2 leaves at $t=100$. SU 3 enters at $t=150$. PU 11 enters at $t=200$. SUs 4--8 enter at $t=250$. We only show PUs 1, 5, 9, 11 (solid
lines) and SUs 1, 2, 3, 4 (dashed lines) in the figure.}\label{fig:Dynamic}
\end{figure}

We demonstrate how the proposed policy can seamlessly adapt to the entry and exit of PUs/SUs. We consider a network with 10 PUs and 2 SUs initially.
The PUs' minimum throughput requirements range from 0.2~bits/s/Hz to 0.38~bits/s/Hz with 0.02~bits/s/Hz increments, namely PU $n$ has a minimum
throughput requirement of $0.2+(n-1)*0.02$~bits/s/Hz. The SUs' have the same minimum throughput requirement of $0.1$~bits/s/Hz. We show the dynamics
of average energy consumptions and throughput of several PUs and all the SUs in Fig.~\ref{fig:Dynamic}.

In the first 100 time slots, we can see that all the users quickly achieve the minimum throughput requirements at around $t=50$. PUs have different
energy consumptions because of their different minimum throughput requirements. The two SUs converge to the same average energy consumption and
average throughput. There are SUs leaving ($t=100$) and entering ($t=150, 250$), and a PU entering ($t=200$). We can see that during the entire
process, the PUs/SUs that are initially in the system maintain the same throughput and energy consumption. The new PU (PU 11) has a higher energy
consumption, because of its higher minimum throughput requirement (0.4~bits/s/Hz), and because of the limited transmission opportunities left for it.
SU 3, however, does not need a higher energy consumption because it occupies the time slots originally assigned to SU 2, who left the network at
$t=100$. But SU 4 does need a higher energy consumption, because there are more SUs and less transmission opportunities in the network after $t=250$.

\section{Conclusion}\label{sec:Conclusion}
In this paper, we proposed nonstationary spectrum sharing policies that allow the PUs and SUs to transmit in a TDMA fashion. The proposed policy can
achieve high spectrum efficiency that is not achievable by existing policies, and is more energy efficient than existing policies under the same
minimum throughput requirements. The proposed policy can achieve high spectrum and energy efficiency even when the users have erroneous and binary
feedback of the interference temperature. We extend the policy to the case with users entering and leaving the network, while still maintaining the
spectrum and energy efficiency of the existing users. The proposed policy is amenable to decentralized implementation and is deviation-proof.
Simulation results demonstrate the significant performance gains over state-of-the-art policies. Interesting future research directions include how
to design the optimal policy when the feedback is finer than binary and when the users have different delay sensitivities (i.e. different discount
factors).

\appendices

\section{Proof of Lemma~\ref{lemma:EqualPowerLevel}}\label{proof:EqualPowerLevel}
Suppose that in the optimal TDMA protocol $\bm{\pi}^*$, there exists a user $i$ and two time slots $t_1 \neq t_2$, such that
$0<\pi_i^*(t_1)<\pi_i^*(t_2)$ (note that we do not assume $t_1<t_2$ or $t_1>t_2$). We will find another protocol $\bm{\pi}^\prime$ that fulfills the
same minimum throughput requirements with lower energy consumptions, which contradicts the fact that $\bm{\pi}^*$ is optimal.

We construct the protocol $\bm{\pi}^\prime$ as follows. The transmission strategies of the users other than user $i$ remain the same, namely
$\bm{\pi}_{-i}^\prime=\bm{\pi}_{-i}^*$. For user $i$, the transmission remains the same for the time slots other than $t_1$ and $t_2$, namely
$\pi_i^\prime(t)=\pi_i^*(t), \forall t\neq t_1,t_2$. Then we increase user $i$'s power level at $t_1$ by $\epsilon_1>0$, i.e.
$\pi_i^\prime(t_1)=\pi_i^*(t_1)+\epsilon_1$, and decrease its power level at $t_2$ by $\epsilon_2>0$, i.e.
$\pi_i^\prime(t_2)=\pi_i^*(t_2)-\epsilon_2$. To maintain user $i$'s average throughput, $\epsilon_1$ and $\epsilon_2$ should satisfy
\begin{eqnarray}
&\!\!\!\!\delta^{t_1}\log_2\big(1+\frac{g_{ii}p_i^*(t_1)}{\sigma_i^2}\big) + \delta^{t_2}\log_2\big(1+\frac{g_{ii}p_i^*(t_2)}{\sigma_i^2}\big) =
\delta^{t_1}\!\log_2\!\big[\!1\!+\!\frac{g_{ii}(p_i^*(t_1)+\epsilon_1)}{\sigma_i^2}\!\big] \!+\!
\delta^{t_2}\!\log_2\!\big[\!1\!+\!\frac{g_{ii}(p_i^*(t_2)-\epsilon_2)}{\sigma_i^2}\!\big]\!. \nonumber
\end{eqnarray}
Given $\epsilon_1$, we can calculate $\epsilon_2$ as $\epsilon_2(\epsilon_1) = \frac{\sigma_i^2+g_{ii} p_i^*(t_2)}{g_{ii}}
\left[1-\left(\frac{\sigma_i^2+g_{ii} p_i^*(t_1)}{\sigma_i^2+g_{ii}(p_i^*(t_1)+\epsilon_1)}\right)^{\delta^{t_1-t_2}}\right]$. Then the decrease in
average energy consumption by switching to protocol $\bm{\pi}^\prime$ can be calculated as $\Delta(\epsilon_1) = -\delta^{t_1} \epsilon_1 +
\delta^{t_2} \epsilon_2(\epsilon_1)$. Taking the derivative of $\Delta(\epsilon_1)$ with respect to $\epsilon_1$, we have
\begin{eqnarray}
&\!\!\!\!\frac{\partial \Delta}{\partial \epsilon_1} = \delta^{t_1}
\left[\!\frac{\sigma_i^2+g_{ii}p_i^*(t_2)}{\sigma_i^2+g_{ii}(p_i^*(t_1)+\epsilon_1)}\left(\frac{\sigma_i^2+g_{ii} p_i^*(t_1)}{\sigma_i^2+g_{ii}
(p_i^*(t_1)+\epsilon_1)}\right)^{\delta^{t_1-t_2}}-1\right]. \nonumber
\end{eqnarray}
Since $p_i^*(t_2)>p_i^*(t_1)$, we have $\frac{\partial \Delta}{\partial \epsilon_1}>0$ when $\epsilon_1=0$. Since $\frac{\partial \Delta}{\partial
\epsilon_1}$ is continuous in $\epsilon_1$ when $\epsilon_1\geq0$, we can find a small enough $\zeta>0$, such that $\frac{\partial \Delta}{\partial
\epsilon_1}>0$ for all $\epsilon_1\in[0,\zeta]$. Hence, the decrease $\Delta(\epsilon_1)$ in user $i$'s average energy consumption by switching to
$\bm{\pi}^\prime$ is positive for any $\epsilon_1\in[0,\zeta]$. This contradicts with the fact that $\bm{\pi}^*$ is optimal, which proves the lemma.

\section{Proof of Theorem~\ref{theorem:CharacterizeEquilibriumPayoff}}\label{proof:CharacterizeEquilibriumPayoff}
Due to space limitation, we present the proof of a simplified version of Theorem~\ref{theorem:CharacterizeEquilibriumPayoff} in the special case when
the users are not self-interested. This proof will illustrate the main idea of the complete proof. Please refer to \cite[Appendix~B]{Appendix} for
the complete proof of Theorem~\ref{theorem:CharacterizeEquilibriumPayoff}.

Specifically, we prove the following lemma on the feasible instantaneous throughput when the users are obedient. The lemma is a special case of
Theorem~\ref{theorem:CharacterizeEquilibriumPayoff} by setting $b_{ij}^+=-\infty$ for all $i,j$.
\begin{lemma}\label{lemma:FeasibleThroughput}
When the users are obedient, an instantaneous throughput vector $\{r_i^{\rm tdma}\}_{i\in\mathcal{M}\cup\mathcal{N}}$ is feasible for the minimum
throughput requirements $\{R_i^{\rm min}\}_{i\in\mathcal{M}\cup\mathcal{N}}$, if
\begin{itemize}
\item the discount factor $\delta$ satisfies $\delta \geq 1-\frac{1}{M+N}$,
\item $\sum_{i\in\mathcal{M}\cup\mathcal{N}} R_i^{\rm min}/r_i^{\rm tdma} = 1$.
\end{itemize}
\end{lemma}
\begin{IEEEproof}
As in dynamic programming, we can decompose each user $i$'s discounted average throughput into the current throughput and the \emph{continuation
throughput} as follows:
\begin{eqnarray}
R_i(\bm{\pi}) &=& (1-\delta) \sum_{t=0}^\infty  \delta^t \cdot (\bm{1}_{\{\pi_i(t)>0\}} \cdot r_i^{\rm tdma}) \nonumber \\
         &=& (1-\delta) \cdot \underbrace{\left(\bm{1}_{\{\pi_i(0)>0\}} \cdot r_i^{\rm tdma}\right)}_{\mathrm{the~current~throughput~at~}t=0} + \delta \cdot \underbrace{\left[(1-\delta) \sum_{t=1}^\infty \delta^{t-1} \cdot (\bm{1}_{\{\pi_i(t)>0\}} \cdot r_i^{\rm tdma})\right]}_{\mathrm{the~continuation~throughput~starting~from~}t=1}. \nonumber
\end{eqnarray}
We can see that the continuation throughput starting from $t=1$ is the discounted average throughput as if the system starts from $t=1$. In general,
we can define user $i$'s continuation throughput starting from $t$ as $\gamma_i(t)\triangleq (1-\delta) \sum_{\tau=t}^\infty \delta^{\tau-t} \cdot
(\bm{1}_{\{\pi_i(\tau)>0\}} \cdot r_i^{\rm tdma})$. Then the decomposition at time $t$ can be written as $\gamma_i(t)=(1-\delta) \cdot
(\bm{1}_{\{\pi_i(t)>0\}} \cdot r_i^{\rm tdma}) + \delta\cdot \gamma_i(t+1)$. Write the continuation throughput vector as
$\bm{\gamma}=(\gamma_1,\ldots,\gamma_N)$.

\begin{definition}[Self-generating set]
A set of throughput vectors $\mathcal{R}$ is a self-generating set, if for any throughput vector $\bm{\gamma}\in\mathcal{R}$, there exists a
$i^*\in\mathcal{N}$ and a continuation throughput vector $\bm{\gamma}^\prime\in\mathcal{R}$ such that for all $i\in\mathcal{N}$,
\begin{eqnarray}
&\gamma_i = (1-\delta)\cdot(\bm{1}_{\{i=i^*\}} \cdot r_i^{\rm tdma}) + \delta \cdot \gamma_{i}^\prime.
\end{eqnarray}
\end{definition}
An important property of the self-generating set, proved in \cite{APS}, is that any throughput vector in $\mathcal{R}$ can be achieved by a TDMA
protocol. This is because for any throughput vector $\bm{\gamma}\in\mathcal{R}$, we can schedule a user $i^*$ to transmit in the current time slot,
and the resulting continuation throughput vector $\bm{\gamma}^\prime$ starting from the next time slot can be decomposed (by a user to transmit and
the following continuation throughput vector) again. We can do the above decomposition iteratively to determine the transmission schedule.

Consider the following set of throughput vectors $\mathcal{R} = \left\{\bm{\gamma}: \sum_{i\in\mathcal{N}} \frac{\gamma_i}{r_i^{\rm tdma}} = 1,
\gamma_i\geq0, \forall i \right\}$. We derive the condition on the discount factor $\delta$ such that $\mathcal{R}$ is self-generating. For a given
vector $\bm{\gamma}\in\mathcal{R}$, if we let user $i$ to transmit, the continuation throughput vector $\bm{\gamma}^\prime$ is
\begin{eqnarray}\label{eqn:ContinuationThroughputVector}
&\gamma_i^\prime = \frac{\gamma_i}{\delta} - \frac{1-\delta}{\delta} \cdot r_i^{\rm tdma},~\mathrm{and}~\gamma_j^\prime =
\frac{\gamma_j}{\delta},~\forall j\neq i.
\end{eqnarray}
To ensure $\bm{\gamma}^\prime\in \mathcal{R}$, the discount factor must satisfy $\delta\geq 1-\frac{\gamma_i}{r_i^{\rm tdma}}$. Hence, to ensure that
any $\bm{\gamma}\in\mathcal{R}$ can be decomposed, the discount factor must satisfy
\begin{eqnarray}\label{eqn:MinimumDiscountFactor}
&\delta\geq \max_{\bm{\gamma}\in\mathcal{R}} \min_{i\in\mathcal{M}\cup\mathcal{N}} \left\{1-\gamma_i/r_i^{\rm tdma}\right\} = 1-\frac{1}{M+N},
\end{eqnarray}
where the optimal solution is achieved when $\gamma_i=\frac{1}{M+N} r_i^{\rm tdma}, \forall i$.
\end{IEEEproof}

\section{Proof of Theorem~\ref{theorem:ITS}}\label{proof:ITS}
We first convert the optimization problem \eqref{eqn:PolicyDesignProblem_OptimalThroughput} into a convex optimization problem. Defining
$x_i=\frac{1}{r_i^{\rm tdma}}$, the objective function can be rewritten as
\begin{eqnarray}
&E\left(\frac{\sigma_1^2 R_1^{\rm min}}{g_{11}}\cdot(2^{\frac{1}{x_1}}-1)\cdot x_1,\ldots,\frac{\sigma_N^2 R_N^{\rm
min}}{g_{NN}}\cdot(2^{\frac{1}{x_N}}-1)\cdot x_N\right).\nonumber
\end{eqnarray}
Based on our assumption, $E(\cdot)$ is convex and increasing in each argument $\frac{\sigma_i^2 R_i^{\rm min}}{g_{ii}}\cdot(2^{\frac{1}{x_i}}-1)\cdot
x_i$. According to the composition rule \cite[Sec.~3.2.4]{Boyd}, $E(\cdot)$ is a convex function of $(x_1,\ldots,x_N)$ if $\frac{\sigma_i^2 R_i^{\rm
min}}{g_{ii}}\cdot(2^{\frac{1}{x_i}}-1)\cdot x_i$ is convex in $x_i$. The second-order derivative of $(2^{\frac{1}{x_i}}-1)\cdot x_i$ is
\begin{eqnarray}\label{eqn:SecondOrder_x}
&\frac{\partial^2~[(2^{\frac{1}{x_i}}-1)\cdot x_i]}{\partial x_i^2} = \ln2 \cdot \frac{2^{\frac{1}{x_i}}}{x_i^3}>0,~\forall x_i>0.
\end{eqnarray}
Hence, the objective function is a convex function of $(x_1,\ldots,x_N)$. It is not difficult to see that the constraints in
\eqref{eqn:PolicyDesignProblem_OptimalThroughput} can be rewritten as linear constraints $\sum_{i\in\mathcal{N}} R_i^{\rm min} \cdot x_i = 1$ and
$x_i\geq\frac{1}{\bar{r}_i}$. As a result, the following optimization problem with decision variables $(x_1,\ldots,x_N)$
\begin{eqnarray}\label{eqn:PolicyDesignProblem_OptimalThroughput_x}
&\displaystyle\min_{(x_1,\ldots,x_N)}& E\left(\left\{\frac{\sigma_i^2 R_i^{\rm min}}{g_{ii}}\cdot(2^{\frac{1}{x_i}}-1)\cdot
x_i\right\}_{i\in\mathcal{N}}\right) \\
&s.t.& \sum_{i\in\mathcal{N}} R_i^{\rm min} \cdot x_i = 1,~x_i\geq\frac{1}{\bar{r}_i},~\forall i\in\mathcal{N}, \nonumber
\end{eqnarray}
is a convex optimization problem.

We solve \eqref{eqn:PolicyDesignProblem_OptimalThroughput_x} by looking at the KKT conditions. Write $\lambda$ as the Lagrangian multiplier of the
constraint $\sum_{i\in\mathcal{N}} R_i^{\rm min} \cdot x_i = 1$, and $\mu_i\geq0$ as the Lagrangian multiplier of the inequality
$x_i\geq\frac{1}{\bar{r}_i}$. The optimal $(x_1^*,\ldots,x_N^*)$ and the optimal $\lambda^*$ and $\mu_i^*$ should satisfy the KKT conditions:
\begin{eqnarray}\label{eqn:KKT_firstorder}
&\frac{\partial E}{\partial x_i}|_{x_i=x_i^*} - \mu_i^* = -\lambda^* R_i^{\min}
\end{eqnarray}
with $\mu_i^*=0$ when $x_i^*>\frac{1}{\bar{r}_i}$, due to the complementary slackness condition. Hence, the problem
\eqref{eqn:PolicyDesignProblem_OptimalThroughput_x} can be solved by finding the optimal $\lambda^*$, such that the solutions $(x_1^*,\ldots,x_N^*)$
to the equations \eqref{eqn:KKT_firstorder} satisfy the equality $\sum_{i\in\mathcal{N}} R_i^{\rm min} \cdot x_i = 1$. Equivalently, we can find the
optimal $\lambda^*$ such that the optimal instantaneous throughput $(r_1^*,\ldots,r_N^*)$ satisfy
\begin{eqnarray}\label{eqn:FirstOrder_r}
&\!\!\!\!\!\!\!\!\frac{\partial E}{\partial x_i^*}|_{x_i^*=\frac{1}{r_i^*}} - \mu_i^* = -\lambda^*
R_i^{\min},~\mathrm{with}~\mu_i^*=0~\mathrm{if}~r_i^*<\bar{r}_i,
\end{eqnarray}
and $\sum_{i\in\mathcal{N}} R_i^{\rm min}/r_i^* = 1$.

Since the first-order derivative $\frac{\partial E}{\partial x_i^*}$ is monotone in $x_i$ (because the second-order derivative is always positive),
we can find the optimal $\lambda^*$ using the bisection method, which converges linearly with rate $\frac{1}{2}$.

\section{Proof of Theorem~\ref{theorem:EquilibriumStrategy}}\label{proof:EquilibriumStrategy}
Due to space limitation, we present the proof of a simplified version of Theorem~\ref{theorem:EquilibriumStrategy} in the special case when the users
are not self-interested. Please refer to \cite[Appendix~C]{Appendix} for the complete proof of Theorem~\ref{theorem:EquilibriumStrategy}.

This proof is closely related to the proof of Theorem~\ref{theorem:CharacterizeEquilibriumPayoff}. Recall that for each continuation throughput
vector $\bm{\gamma}(t)$ at time $t$, if we choose user $i$ to transmit, we can calculate the resulting continuation throughput vector
$\bm{\gamma}(t+1)$ at time $t+1$ as in \eqref{eqn:ContinuationThroughputVector}. The proof ofTheorem~\ref{theorem:CharacterizeEquilibriumPayoff}
ensures that as long as we choose the user to transmit at time $t$ based on $i=\arg\min_{j\in\mathcal{M}\cup\mathcal{N}}
\left\{1-\gamma_j(t)/r_i^*\right\}$ (see \eqref{eqn:MinimumDiscountFactor}), the continuation throughput vector $\bm{\gamma}(t+1)$ at time $t+1$ will
also be achievable. The LDF scheduling schedules the transmission exactly in this way in each time slot. By setting the continuation throughput at
time $0$ as $\gamma_i(0)=R_i^{\rm min}$, each user $i$ can achieve the average throughput $R_i^{\rm min}$. Since the instantaneous throughput is the
optimal one, $r_i^*$, the energy efficiency criterion is minimized.

Note that $R_i^{\rm min}=(1-\delta) \sum_{\tau=0}^\infty \delta^\tau \cdot (\bm{1}_{\pi_i(\tau)>0} \cdot r_i^*) = (1-\delta) \sum_{\tau=0}^t
\delta^\tau \cdot (\bm{1}_{\pi_i(\tau)>0} \cdot r_i^*) + (1-\delta) \sum_{\tau=t+1}^\infty \delta^\tau \cdot (\bm{1}_{\pi_i(\tau)>0} \cdot r_i^*)$.
Since $0\leq(1-\delta) \sum_{\tau=t+1}^\infty \delta^\tau \cdot (\bm{1}_{\pi_i(\tau)>0} \cdot r_i^*)\leq (1-\delta) \sum_{\tau=t+1}^\infty
\delta^\tau \cdot r_i^* = \delta^{t+1}\cdot r_i^*$, we have $|(1-\delta) \sum_{\tau=0}^t \delta^\tau \cdot (\bm{1}_{\pi_i(\tau)>0} \cdot r_i^*) -
R_i^{\rm min}| \leq r_i^*\cdot\delta^{t+1}$.

\section{Proof of Theorem~\ref{theorem:DynamicEntryExit}}\label{proof:DynamicEntryExit}
For a user $i$, consider the distance between its average throughput at time $t$ and its minimum throughput $R_i^{\rm min}$. Suppose that each time
slot $\tau$ is in the $k_\tau$th epoch (time slot $t$ is in the $\ell$th epoch), and that the beginning of the $k$th epoch is $t_k$ with $t_0=0$.
Then the distance is
\begin{eqnarray}
& & \begin{array}{l}\left|(1-\delta) \sum_{\tau=0}^{t} \delta^\tau (\bm{1}_{\pi_i(\tau)>0}\cdot r_i^{(k_\tau)}) - R_i^{\rm min}\right|\end{array} \nonumber\\
&=& \begin{array}{l}\left|\left[(1-\delta) \sum_{\tau=t_0}^{t_1-1} \delta^\tau (\bm{1}_{\pi_i(\tau)>0}\cdot r_i^{(0)}) - R_i^{\rm min}\right] + (1-\delta) \sum_{\tau=t_1}^{t} \delta^\tau (\bm{1}_{\pi_i(\tau)>0}\cdot r_i^{(k_\tau)})\right|\end{array} \nonumber \\
&=& \begin{array}{l}\left|(1-\delta) \sum_{\tau=t_1}^{t} \delta^\tau (\bm{1}_{\pi_i(\tau)>0}\cdot r_i^{(k_\tau)}) - \gamma_i(t_1)\right|\end{array} \nonumber \\
&=& \begin{array}{l}\left|(1-\delta) \sum_{\tau=t_{\ell}}^{t} \delta^\tau (\bm{1}_{\pi_i(\tau)>0}\cdot r_i^{(\ell)}) -
\gamma_i(t_{\ell})\right|\end{array} \nonumber.
\end{eqnarray}
Since $\gamma_i(t_{\ell})$ is the input to the LDF scheduling at the beginning of the $\ell$th epoch, from Theorem~\ref{theorem:EquilibriumStrategy},
we have $\left|(1-\delta) \sum_{\tau=t_{\ell}}^{t} \delta^\tau (\bm{1}_{\pi_i(\tau)>0}\cdot r_i^{(\ell)}) - \gamma_i(t_{\ell})\right|\leq
r_i^{(\ell)} \cdot \delta^{t+1}$. Hence, the distance between the average throughput and the minimum throughput requirement decreases exponentially
with time even with users entering and leaving.

\end{document}